\documentstyle[pre,multicol,aps]{revtex}

\begin{document}

\input{epsf.tex}
\epsfclipon

\title{ Nonlinearity and disorder:  
        Classification and stability of nonlinear impurity modes }
\author{Andrey A. Sukhorukov$^1$, Yuri S. Kivshar$^1$, 
Ole Bang$^{1,2}$, Jens J. Rasmussen$^3$, and Peter L. Christiansen$^2$}

\address{
$^1$  Optical Sciences Centre,  Australian National University,  
      Canberra ACT 0200, Australia  \\
$^2$  Department of Mathematical Modelling, 
      Technical University of Denmark, 
      bldg. 321, 2800 Kgs. Lyngby, Denmark \\
$^3$  Ris{\o} National Laboratory, Optics and Fluid Dynamics
      Department, Ris{\o}, Roskilde DK-4000, Denmark }

\maketitle

\begin{abstract}
We study the effects produced by competition of two physical mechanisms of energy localization in {\em inhomogeneous nonlinear} systems. As an example, we analyze spatially localized modes supported by a nonlinear impurity in the generalized nonlinear Schr\"odinger equation and describe {\em three types} of nonlinear impurity modes~--- one- and two-hump symmetric localized modes and asymmetric localized modes~--- for both focusing and defocusing nonlinearity and two different (attractive or repulsive) types of impurity. We obtain {\em an analytical stability criterion} for the nonlinear localized modes and consider the case of a power-law nonlinearity in detail. We discuss several scenarios of the instability-induced dynamics of the nonlinear impurity modes, including the mode decay or switching to a new stable state, and {\em collapse at the impurity site}.
\end{abstract}


\vspace*{-1.0 cm}

\begin{multicols}{2}

\narrowtext

\section{Introduction}

Wave scattering by localized impurities (or defects) is a fundamental problem of solid state physics~\cite{book}. Impurities break the translational symmetry of a physical system and lead to several effects such as wave reflection, resonant scattering, and excitation of {\em impurity modes}~--- spatially localized oscillatory states at the impurity sites~\cite{lifshitz}. These two kinds of problems, i.e. wave scattering in inhomogeneous media and defect-supported localized modes, appear in many different physical problems, such as the scattering of surface acoustic waves by surface defects or interfaces~\cite{acoustic}, excitation of defect modes in superconductors in the vicinity of the twinning planes~\cite{super} and high-$T_c$ superconductors~\cite{highTc}, the dynamics of the tight-binding Holstein-type models of the electron-phonon coupling~\cite{molina,molina2}, light propagation in dielectric super-lattices with embedded defect layers~\cite{souk}, excitation of defect states in photonic crystal waveguides~\cite{pc}, light trapping and switching in non-uniform waveguide arrays~\cite{arrays}, etc. In all such cases, the impurities (or defects) lead to the energy trapping and localization in the vicinity of the defects, that occurs in the form of {\em spatially localized impurity modes}.

When nonlinearity becomes important, it may lead to self-trapping and energy localization even in a perfect (or homogeneous) system in the form of {\em intrinsic localized modes}. Spatially localized modes of nonlinear systems are usually associated with {\em solitary waves} (or solitons) in continuous models, or {\em discrete breathers} in lattice models; and they have been a subject of intensive studies during the past years~\cite{flach}. However, the study of nonlinear phenomena in inhomogeneous and disordered systems is {\em still largely an open area of research}~\cite{review}. Simultaneous presence of nonlinearity and disorder is associated with various dynamical processes in solids, biological systems, and optics~\cite{books}. For example, nonlinearities can become important even in a harmonic lattice due to the interaction of an exciton with the lattice vibrations~\cite{tsir}, where impurities may appear as a result of doping of materials with atoms or molecules that have stronger local coupling. Such impurities can also appear in (inherently nonlinear) spin-wave systems due to, e.g., a local variation of the coupling between neighboring spins~\cite{magnet}.

When both nonlinearity and disorder are present simultaneously, it is expected that competition between two different mechanisms of energy localization (i.e. one, due to the self-action of nonlinearity, and the other one, due to localization induced by disorder) will lead to a complicated and somewhat nontrivial physical picture of localized states and their stability. In this paper, we consider one of the examples of such a competition, and analyze different types of nonlinear localized modes and their stability in the framework of the generalized nonlinear Schr\"odinger (NLS) equation with a point-like impurity. 

The problem we analyze here has a number of important physical applications ranging from the nonlinear dynamics of solids~\cite{super,molina,tsir,magnet} to the theory of nonlinear photonic crystals~\cite{souk,pc} and waveguide arrays~\cite{arrays} in optics. In application to the theory of electromagnetic waves, this problem describes a special case of a stratified (or layered) dielectric medium for which nonlinear guided waves and their stability has been analyzed during the last 20 years~\cite{optics,ms}. For other applications, the theory of nonlinear localized modes is less developed and, in particular, only a few publications~\cite{molina2,malomed,nls_imp,ltp} addressed the important issue of stability of nonlinear impurity modes. In this paper, we study the properties of nonlinear impurity modes in the framework of the generalized NLS equation and develop, for the first time to our knowledge, {\em a systematic classification and linear stability analysis} of spatially localized impurity modes of three distinct types. Our results can be linked to different special cases of the theory of nonlinear guided waves in layered dielectric media, and they also provide a generalization of the theory of nonlinear impurity modes in solids, together with the analysis of their stability and instability-induced dynamics, emphasizing the cases where we can observe a clear evidence of competition between the two physical mechanisms of energy localization.

The paper is organized as follows. In Sec.~\ref{sect:model} we discuss our physical model and describe three different types of nonlinear localized modes supported by a point-like impurity: symmetric one- and two-hump modes and asymmetric modes. Section~\ref{sect:stab} includes a summary of the results of the linear stability analysis, and it also presents the analytical criteria of the mode stability, for a general form of the NLS equation with a nonlinear impurity. A detailed analysis of the mode structure and stability, as well as the discussion of the competition between two different mechanisms of energy localization, are presented in Secs.~\mbox{\ref{sect:power_nl}-\ref{sect:pw_defoc}}, for the particular case of the power-law nonlinearity and both attractive and repulsive impurities. At last, Sec.~\ref{sect:collapse} discusses two different types of the nonlinearity-induced collapse dynamics of the nonlinear modes, including collapse at the impurity site.

\section{Model and localized modes} 
         \label{sect:model} \label{sect:stationar}

We consider a general problem in which the dynamics of elementary excitations of a physical system (e.g., phonon, magnons, etc.) is described by an effective equation for the wave-packet envelope $\psi (x,t)$~\cite{kosevich}. When the density of such quasi-particles becomes high enough, their interaction should be taken into account, e.g. in the framework of the mean-field approximation. In the simplest case, the quasi-particle interaction and collective phenomena in an inhomogeneous medium can be described by the nonlinear Schr\"odinger (NLS) equation for the wave-packet envelope $\psi (x,t)$,
\begin{equation} \label{eq:nls}
     i \frac{\partial \psi}{\partial t} 
     + \frac{\partial^2 \psi}{\partial x^2}  
     + {\cal F}(I; x) \psi = 0,
\end{equation}
where $I \equiv |\psi|^2$ characterizes the density of the quasi-particles, $t$ is time, $x$ is the spatial coordinate, and the real function ${\cal F}(I; x)$ describes both {\em nonlinear} and {\em disordered} properties of the medium. We consider the case when the inhomogeneity is localized in a small region. Then, if the corresponding wavelength is much larger than the defect size, in the continuum limit approximation the inhomogeneity can be modeled by a delta-function and, therefore, we can write
\begin{equation} \label{eq:def_F}
      {\cal F}(I; x) = F(I) + \delta(x) G(I), 
\end{equation}
where the functions $F(I)$ and $G(I)$ characterise the properties of the bulk medium and impurity, respectively. Hereafter we assume that the nonlinear term $F(I)$ does not include a constant linear part, i.e. $F(0) = 0$, as otherwise it is always possible to re-scale the original Eq.~(\ref{eq:nls}) by introducing a new function $\widetilde{\psi} = \psi \exp[- i F(0) t]$. 

The model~(\ref{eq:nls}),(\ref{eq:def_F}) appears in different physical problems of the macroscopic nonlinear dynamics of solids and nonlinear optical systems. In particular, it describes a special case of a more general problem of the existence and stability of nonlinear guided waves in a layered (or stratified) dielectric medium, where the delta-function defect corresponds to a very thin layer embedded into a nonlinear medium with a Kerr or non-Kerr response (see, e.g., Ref.~\cite{optics}); in this case the time variable $t$ stands for the propagation coordinate along the layer, and $x$ is the transverse coordinate.

For spatially localized solutions, Eqs.~(\ref{eq:nls})-(\ref{eq:def_F}) conserve the power,
\begin{equation} \label{eq:P}
   P = \int_{-\infty}^{+\infty} |\psi(x)|^2 dx  ,
\end{equation}
and the Hamiltonian
\begin{equation} \label{eq:H}
    H = \int_{-\infty}^{+\infty} \left\{
           {\left| \frac{\partial \psi}{\partial x} \right|}^2 
           - \int_0^{I(x)} 
                {\cal F}(I^{\prime}; x)  dI^{\prime} 
      \right\} dx  .
\end{equation}
Note, however, that the total momentum is not conserved since the 
translational invariance of the model~(\ref{eq:nls})-(\ref{eq:def_F}) 
is broken by the presence of the inhomogeneity. 

We look for spatially localized stationary solutions of Eqs.~(\ref{eq:nls})-(\ref{eq:def_F}) in the standard form, 
\[
  \psi(x,t) = u(x) e^{i \omega t},
\]
where $\omega>0$ is the mode frequency ($\omega>0$ for the mode to be 
exponentially localized), and the real function $u(x)$ satisfies 
the equation:
\begin{equation} \label{eq:u}
  - \omega u + \frac{d^2 u}{d x^2} + F(I) u +  \delta(x) G(I) u = 0.
\end{equation}

Let us first discuss the well-known case when the impurity is absent,  $G(I) = 0$. Then, a localized solution $u_0(x)$ of the reduced Eq.~(\ref{eq:u}),
\begin{equation} \label{eq:u0}
  - \omega u_0 + \frac{d^2 u_0}{d x^2} + F(u_0^2) u_0 = 0,
\end{equation}
describes a self-trapped state in a uniform nonlinear medium. Due to the translational symmetry, a localized solution of Eq.~(\ref{eq:u0}) can be presented in the form $u_0(x - x_0)$, where $x_0$ is {\em an arbitrary position}. We also note that the mode profile is symmetric, $u_0(x) = u_0(-x)$, it does not contain zeros, $u_0(x)>0$, and it has a single hump since $d u_0 / d |x| < 0$ at $x \ne 0$. Such a localized state is possible in a {\em self-focusing medium} in the form of a bright solitary wave (or soliton), which is characterized by the power
\begin{equation} \label{eq:P0}
   P_0( \omega ) = 2\int_0^\infty u_0^2(x)\; d x .
\end{equation}
In contrast, in a {\em defocusing medium}, spatially localized solutions of Eq.~(\ref{eq:u0}) do not exist. However, certain {\em singular solutions} with zero asymptotic at $x \rightarrow \pm \infty$ are still possible in this case, and they will play an important role for constructing localized impurity modes.

When a point-like defect is introduced into the system, the translational invariance of the model is broken at the defect location, $x=0$. Nevertheless, the nonlinear modes of the inhomogeneous model~(\ref{eq:u}) localized near the impurity site $x=0$ can be constructed by using the solutions of the homogeneous equation~(\ref{eq:u0}). Indeed, such a solution should satisfy Eq.~(\ref{eq:u0}) on both sides of the defect, and it can therefore be presented in the following general form,  
\begin{equation} \label{eq:u_slv}
   u(x) = \left\{ 
   \begin{array}{l}
      u_0( x - x_0 )   ,\;\;  x \ge 0 , \\
      u_0( x - s x_0 ) ,\;\;  x \le 0 , 
   \end{array} 
   \right.
\end{equation}
where $x_0$ and $s$ are yet unknown parameters. Our task now is to satisfy 
Eq.~(\ref{eq:u}) at $x = 0$ in order to define the unknown parameters $x_0$ and $s$ through the impurity characteristics. The first constraint follows 
from the field continuity at the impurity site, $u(0^+) = u(0^-)$. 
For Eq.~(\ref{eq:u_slv}) it immediately yields the condition $s = \pm 1$. 
Thus, the parameter $s$ defines the symmetry of the localized mode, i.e. the mode is {\em symmetric} for $s = -1$ and {\em asymmetric} for $s = +1$. 
In order to derive the second matching condition, we integrate Eq.~(\ref{eq:u}) over an infinitely small segment around the impurity point $x = 0$ and obtain  the transcendental equation for the parameter $x_0$:
\begin{equation} \label{eq:x0}
   (1-s) \frac{d u_0}{d x} (x_0) = G(I_0) u_0(x_0),
\end{equation}
where $I_0 = u_0^2(x_0)$.

\begin{figure}
\setlength{\epsfxsize}{8.0cm}
\vspace{2mm}
\centerline{\mbox{\epsffile{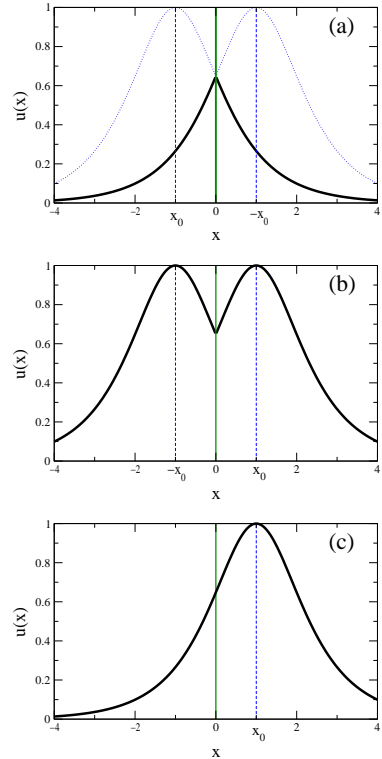}}}
\vspace{2mm}
\caption{ \label{fig:eprf}
Characteristic profiles of nonlinear impurity modes in a self-focusing medium: 
(a,b)~symmetric modes supported by an attractive ($x_0<0$) and repulsive 
($x_0>0$) impurity, respectively; 
(c)~asymmetric impurity mode with $x_0>0$.
}
\end{figure}

Let us discuss the general properties of the impurity modes as localized 
solutions of a homogeneous medium satisfying the matching conditions 
obtained above. {\em Symmetric modes} ($s=-1$) can be presented in the
form $u(x) = u_0(|x| - x_0)$, and have the power
\begin{equation} \label{eq:Pm}
   P( \omega ) = 2 \int_{-x_0}^{+\infty} u_0^2(x)\; d x .
\end{equation}
Using the properties of the homogeneous solitary wave solution $u_0(x)$ we see that if the impurity is {\em attractive} (i.e.~$G(I_0)>0$) then $x_0<0$. Thus the resulting symmetric profile $u(x)$ is single-humped [see Fig.~\ref{fig:eprf}(a)] and $P<P_0$. When the impurity is {\em repulsive} (i.e.~$G(I_0)<0$) then $x_0>0$ and thus the resulting profile is double-humped with $P>P_0$, as shown in Fig.~\ref{fig:eprf}(b). We note that the single-hump solution may still exist in a self-defocusing medium being constructed from two pieces of a singular solution of the homogeneous model (see below), whereas the two-hump solutions are possible only in a self-focusing medium.

For {\em asymmetric modes} $s=+1$ and the left-hand side of Eq.~(\ref{eq:x0}) vanishes. Thus solutions are possible only if the impurity response vanishes as well, i.e.~$G(I_0) = 0$. This condition can be satisfied when the nonlinear and linear parts of the impurity response compensate each other, i.e. for certain values of $I_0$. Remarkably, the profile of the asymmetric impurity mode coincides with that of a bright soliton in an uniform self-focusing medium, $u(x)=u_0(x-x_0)$, and thus $P=P_0$. However, its position is fixed by the impurity according to the relation $G(I_0) = 0$ [see Fig.~\ref{fig:eprf}(c)]. Naturally there always exist two degenerate asymmetric solutions corresponding to positive and negative $x_0$, respectively.

\section{Stability analysis} \label{sect:stab}
\subsection{General formalism} \label{sect:stab_general}

One of the key properties of a nonlinear localized mode is its {\em linear stability} determined by the character of the mode dynamics under the action of small perturbations of its stationary state. In general, two different scenarios of the perturbation-induced mode dynamics are possible. In the first case, a nonlinear mode can acquire only small distortions to its steady-state profile and the parameters of a nonlinear mode oscillate in the vicinity of its stationary state. In this case, we call such a nonlinear mode {\em linearly stable}. On the other hand, under the influence of small perturbations initial deviations of the nonlinear mode parameters from their stationary values can grow exponentially; and in this case we define the nonlinear mode as {\em linearly unstable}.

We consider localized modes that are square- or $L^2$-integrable (have 
finite power or $L^2$-norm), and have profile functions that belong to 
a Hilbert space whose inner product is the $L^2$-norm
\begin{equation} \label{innerproduct}
  \{a(x),b(x)\} \equiv \int_{-\infty}^\infty a^*(x)b(x) \, dx ,
\end{equation}
where $a^*(x)$ denotes the complex conjugate of the function $a(x)$. 
To find the linear stability conditions we consider the evolution of a
small-amplitude perturbation $f(x,t)$ of the stationary solution, i.e.
\begin{equation} \label{eq:psi}
  \psi(x,t) = \left[ u(x) + f(x,t) \right] e^{i \omega t} .
\end{equation}
Writing the perturbation in the form
\begin{equation} \label{eq:emode}
  f(x,t) = \left[ v(x) - w(x) \right] e^{ i \Omega t} 
         + \left[ v^{\ast}(x) + w^{\ast}(x) \right] 
                                      e^{ -  i \Omega^{\ast} t } ,
\end{equation}
and substituting Eq.~(\ref{eq:psi}) into Eq.~(\ref{eq:nls}), we obtain
an eigenvalue problem for the functions $v(x)$ and $w(x)$, 
\begin{equation} \label{eq:L_vw}\label{eq:Lj}
 \begin{array}{l} {\displaystyle
   L_0 w = \Omega v, \quad L_1v=\Omega w,} 
 \end{array}
\end{equation}
where $\Omega$ is the complex spectral eigenvalue, and
\[
 \begin{array}{l} {\displaystyle
    L_j=-d^2/dx^2+\omega-U_j, \quad U_j=F_j+G_j\delta(x)
 } \\*[9pt] {\displaystyle
    F_0 = F(u^2), \quad F_1 = F_0 +  2 u^2 F'(u^2),
 } \\*[9pt] {\displaystyle
    G_0 = G(I_0), \quad
    G_1 = G_0 + 2 I_0 G'(I_0).
 } \end{array}
\]
Here prime denotes differentiation with respect to the argument.
It follows from Eq.~(\ref{eq:emode}) that the mode is stable if 
all the eigenvalues $\Omega$ are real and unstable otherwise.

To proceed we reduce Eqs.~(\ref{eq:L_vw}) to a single equation:
\begin{equation} \label{eq:L_v}
  L_0 L_1 v = \Omega^2 v , 
\end{equation}
where stability requires all eigenvalues $\Omega^2$ to be positive. 
It is straightforward to show that $L_0=L^{+} L^{-}$, where $L^\pm = 
\pm d / dx\; + u^{-1} (du/dx)$, and thus one can instead consider the 
auxiliary eigenvalue problem,
\begin{equation} \label{eq:L_vt}
  L^{-} L_1 L^{+} \tilde{v} = \Omega^2 \tilde{v} , \nonumber
\end{equation}
which reduces to Eq.~(\ref{eq:L_v}) after the substitution $v=L^{+}\tilde{v}$. Because the operator $L^{+}$ has only a single neutral mode, $v(x)=u^{-1}(x)$, which is not an eigenmode of Eqs.~(\ref{eq:L_v}-\ref{eq:L_vt}), these two eigenvalue problems have equivalent spectra (see Ref.~\cite{ltp}). Since the operator $L^{-} L_1 L^{+}$ is Hermitian all eigenvalues $\Omega^2$ of Eqs.~(\ref{eq:L_v}) and (\ref{eq:L_vt}) are real.

The operators $L_j$ ($j = 1,2$) are well-studied in the literature, in particular as a characteristic example of the spectral theory of second-order differential operators (see Ref.~\cite{titchmarsh}). For our problem we use two {\em general mathematical results} about the spectrum of the linear eigenvalue problem $L_j \varphi_n^{(j)} = \lambda_n^{(j)} \varphi_n^{(j)}$: (i) the eigenvalues can be ordered as $\lambda_{n+1}^{(j)} > \lambda_n^{(j)}$ where $n \ge 0$ defines the number of zeros in the corresponding eigenfunction $\varphi_n^{(j)}$; (ii) for ``deeper'' potential wells, $\widetilde{U}_j(x) \ge U_j(x)$, the corresponding set of eigenvalues is shifted ``down', $\widetilde{\lambda}_n^{(j)} \le \lambda_n^{(j)}$.

Let us fist discuss the properties of the operator $L_0$ whose spectrum we define as $\lambda_n^{(0)}$. The neutral mode of $L_0$ is the stationary solution of Eq.~(\ref{eq:u}), since $L_0 u(x) = 0$. Moreover, $u(x)>0$ is the ground state solution with no nodes and therefore $\lambda_n^{(0)} > \lambda_0^{(0)} = 0$ for $n>0$. This means that the operator $L_0$ is positive definite on the subspace of functions orthogonal to $u(x)$, which allows to use several general theorems~\cite{ms,vk,kuznetsov,jones,grillakis} to link the stability properties to the number of negative eigenvalues of the operator $L_1$, whose spectrum we define as $\lambda_n^{(1)}$: (i)~the mode is unstable if there are two (or more) negative eigenvalues, i.e.~if $\lambda_1^{(1)} < 0$; (ii)~the mode is stable if $L_1$ is positively definite, i.e.~if $\lambda_0^{(1)} > 0$; (iii)~in the intermediate case the stability depends on the slope of the function $P(\omega)$ according to the Vakhitov-Kolokolov~(VK) criterion~\cite{vk}, i.e.~the mode is stable if $dP/d\omega > 0$ and unstable otherwise. Thus, to distinguish between these cases, it is sufficient to determine the signs of $\lambda_0^{(1)}$ and $\lambda_1^{(1)}$. 

In general, the spectral properties of the linear operator $L_1$ depend  on the mode frequency $\omega$, and the number of its negative eigenvalues  can change. This is associated with the so-called {\em critical points} in  the power dependence $P(\omega)$ for the localized mode~\cite{ms}.  One type of such critical points is a {\em bifurcation} into two families  of solutions, i.e.~symmetric and asymmetric ones, which exist for the same mode frequency.
Thus, the stability properties of these distinct types of nonlinear  localized modes can be expected to be different, so we discuss them  separately.

\subsection{Symmetric modes} \label{sect:stab_symm}

For symmetric modes with $s=-1$ the operator $L_1$ has the symmetric
neutral mode with zero eigenvalue
\[
  L_1(du/d|x|) = 0 , 
\]
for the special value of the defect response
\begin{equation} \label{eq:M1_cr}
  G_1^{\rm cr} = \left.  2 \left( \frac{d u_0(x)}{dx} \right)^{-1} 
                      \frac{d^2 u_0(x)}{dx^2} \right|_{x=x_0} .
\end{equation}
For $x_0>0$ the profile $u(x)$ has two humps and the neutral mode corresponds therefore to the second eigenmode $\varphi_2^{(1)}$ (two nodes) with $\lambda_2^{(1)}=0$. Thus the two-hump symmetric modes are unstable (two negative eigenvalues). For $x_0<0$ the neutral mode is the nodeless ground-state $\varphi_0^{(1)}$ with $\lambda_0^{(1)}=0$. Thus the single-hump symmetric modes are stable (no negative eigenvalues), if $G_1 = G_1^{\rm cr}$.

For other values of the defect response $G_1$ it follows from the spectral theorem that the eigenvalue of the neutral mode decreases when $G_1>G_1^{ \rm cr}$ (deeper well) and increases when $G_1<G_1^{\rm cr}$. Therefore, for $x_0 < 0$ and $G_1 < G_1^{\rm cr}$ we have $\lambda_0^{(1)} > 0$ and the modes are stable. On the other hand, due to the symmetry of the potentials in the linear eigenvalue problem, $U_j(x) = U_j(-x)$, the amplitude of the first-order eigenmode vanishes at the defect location ($x=0$), and thus its eigenvalue $\lambda_1^{(1)}$ does not depend on $G_1$. However, the inequality $\lambda_0^{(1)} < \lambda_1^{(1)} < \lambda_2^{(1)}$ is always fulfilled for any $G_1$, and thus $\lambda_1^{(1)}<0$ if $x_0>0$ and $\lambda_1^{(1)}>0$ if $x_0<0$.  

In the special case when $x_0=0$ it is straightforward to show that the function $du(x)/d x$, which has a single zero at $x=0$, is a neutral mode of $L_1$ with $\lambda_1^{(1)}=0$. In Table~\ref{tab:stab_symm} we summarize the properties of the operator $L_1$ and the corresponding general conditions for the stability of nonlinear localized modes.

\noindent \parbox[l]{8cm}{
\noindent
\begin{table}
\caption{Stability conditions for symmetric modes}
\begin{tabular}{lll} 
     Condition
   & $L_1$ spectrum
   & Stability 
\\ \hline \hline
     $x_0 > 0$ 
   & $\lambda_0^{(1)} < \lambda_1^{(1)} < 0$
   & unstable
\\ \hline
     $x_0 \le 0$, $G_1 \le G_1^{\rm cr}$
   & $0 \le \lambda_0^{(1)} < \lambda_1^{(1)}$
   & stable
\\ \hline 
     $x_0 \le 0$, $G_1 > G_1^{\rm cr}$, $dP/d\omega > 0$
   & $\lambda_0^{(1)} < 0 \le \lambda_1^{(1)}$
   & stable
\\ \hline 
     $x_0 \le 0$, $G_1 > G_1^{\rm cr}$, $dP/d\omega \le 0$
   & $\lambda_0^{(1)} < 0 \le \lambda_1^{(1)}$
   & unstable
\end{tabular}
\label{tab:stab_symm}
\end{table}
}

It is important to connect the spectral characteristics of the operator $L_1$, and thus the stability properties, with the character of the power functional $P(\omega)$. To do so we notice that at the {\em bifurcation point}, defined as $P(\omega) = P_0(\omega)$, the parameter $x_0$ and thus the eigenvalue $\lambda_1^{(1)}$ changes sign. The two-hump symmetric modes with $P(\omega) > P_0(\omega)$, for which $x_0>0$ and thus $\lambda_1^{(1)}<0$, are therefore {\em always unstable}. Instead a new family of asymmetric modes emerges at the bifurcation point, whose stability properties we discuss in the next section.

The single-hump symmetric modes (i.e.~$x_0<0$) can only change their stability properties through a change of sign of the lowest eigenvalue $\lambda_0^{(1)}$ or the slope $dP/d\omega$, according to Table~\ref{tab:stab_symm}. In order to find the critical points associated with the transition of $\lambda_0^{(1)}$ through zero we employ the approach developed in Ref.~\cite{ms} and differentiate Eq.~(\ref{eq:u}) with respect to the mode power $P$, treating $\omega$ as a function of $P$. This gives us the relation
\[
  u(x)(\partial \omega / \partial P) = -L_1(\partial u / \partial P) .
\]
The points where $d\omega/dP = 0$ or $u \equiv 0$ are thus critical points, since there the operator $L_1$ has a zero eigenvalue $\lambda_0^{(1)} = 0$ with eigenfunction $\partial u/\partial P$. It is possible to show that these are the only critical points~\cite{ms}. Moreover, it can be demonstrated that $G_1 > G_1^{\rm cr}$ for the branch originating from the critical point with the positive slope, $d P / d \omega > 0$, while $G_1 < G_1^{\rm cr}$ if $d P / d \omega < 0$ in the vicinity of the critical point. Therefore, the localized modes are always stable close to such critical points. This conclusion implies, in particular, stability of modes with a vanishing power, $P \rightarrow 0$. In this linear limit the impurity can support a localized mode only when it is attractive (i.e.~$G_0 > 0$), and the mode frequency is $\omega_0 = G_0^2/4$.

\subsection{Asymmetric modes} \label{sect:stab_asymm}

The profiles of asymmetric nonlinear modes coincide with those of solitons (since $G_0 = 0$), but the spectrum of the operator $L_1$ can become different. Only in the special case $G_1 = 0$, there exists a first-order (one node) neutral mode with $\lambda_1^{(1)} = 0$, i.e.~$L_1(du/dx) = 0$. Then, from the spectral theorems~\cite{titchmarsh} it follows that $\lambda_1^{(1)}<0$ if $G_1>0$ and $\lambda_1^{(1)}>0$ if $G_1<0$. Thus the asymmetric mode is always unstable for $G_1>0$, with respect to translational shifts along the $x$ axis.

To determine the sign of $\lambda_0^{(1)}$ for $G_1<0$ we consider, 
without a lack of generality, the case $x_0 > 0$ and study the 
operator $\widetilde{L}_1$ with the potential
\begin{equation} \label{eq:L1_inf}
  \widetilde{U}_1(x) = \left\{ \begin{array}{l}
                  F_1\left[ u_0^2(x-x_0) \right],\; x>x_0 , \\
                  -\infty,\; x < x_0.
         \end{array} \right.
\end{equation}
It is straightforward to check that the lowest eigenvalue of $\widetilde{L}_1$ is zero with the ground state
\[
  \tilde{\varphi}_0^{(1)} = \left\{ \begin{array}{l}
                  d u_0(x - x_0) / d x ,\; x>x_0 , \\
                  0 ,\; x < x_0.
         \end{array} \right.
\]
For $G_1<0$ we have $U_1(x) > \widetilde{U}_1(x)$, and thus according to the spectral theorem the eigenvalues of the corresponding operators are related as $\lambda_n^{(1)} < \tilde{\lambda}_n^{(1)}$. Therefore, $\lambda_0^{(1)}$ is always negative, meaning that the VK theorem applies. We combine these findings with the general stability criteria~\cite{ms,vk,kuznetsov,jones,grillakis}, and summarize the stability conditions in Table~\ref{tab:stab_asymm}.

\noindent \parbox[l]{8cm}{
\noindent
\begin{table}
\caption{ Stability conditions for asymmetric modes }
\begin{tabular}{lll}
     Condition
   & $L_1$ spectrum
   & Stability
\\ \hline \hline
     $G_1 > 0$
   & $\lambda_0^{(1)} < \lambda_1^{(1)} < 0$
   & unstable
\\ \hline
     $G_1 \le 0$, $dP_0/d\omega > 0$
   & $\lambda_0^{(1)} < 0 \le \lambda_1^{(1)}$
   & stable
\\ \hline
     $G_1 \le 0$, $dP_0/d\omega \le 0$
   & $\lambda_0^{(1)} < 0 \le \lambda_1^{(1)}$
   & unstable
\end{tabular}
\label{tab:stab_asymm}
\end{table}
}

We note that, for a given defect response $G(I)$, there can exist several families of asymmetric modes each characterized by the intensity at the defect $I_0$, satisfying the relation $G(I_0) = 0$. There are always at least two solutions with the same $I_0$ but $\pm x_0$. All these degenerate families have the same power and are thus indistinguishable on the $P(\omega)$ diagram. To determine the stability we look at the bifurcation from the symmetric modes at $P(\omega)=P_0(\omega)$ (or $x_0 = 0$). Performing the analysis similar to that in Ref.~\cite{ms}, we find that after the bifurcation $G_1<0$ if the branch for symmetric modes is above that of the asymmetric modes, $P(\omega)>P_0(\omega)$. In the opposite case $G_1>0$ and the asymmetric modes are always unstable.

\section{Power-law nonlinearity} \label{sect:power_nl}

We now demonstrate the characteristic existence and stability features 
of nonlinear localized impurity modes, by applying the results of 
Secs.~\ref{sect:stationar} and \ref{sect:stab} to the illustrative 
case of power-law nonlinearities.
In this Section we consider the analytically obtainable results.
The detailed numerical results are presented in Secs.~\ref{sect:pw_foc}
and \ref{sect:pw_defoc}.

\subsection{Solitons in homogeneous media}
            \label{sect:power_nl_sol}

To construct the general solutions we consider first the special case of
a homogeneous medium with $G\equiv 0$ and the bulk power-law nonlinearity
\begin{equation} \label{eq:NG_power_F}
   F(I) = \rho I^{\sigma}, \quad \sigma>0.
\end{equation}
For self-focusing bulk media $\rho=+1$ and for defocusing bulk media
$\rho=-1$.  The profiles of the spatially localized stationary solutions 
can be found in the form
\begin{equation} \label{eq:u0_power}
   u_0(x) = A_0 \left\{ 
   \begin{array}{l}
      \cosh^{-1 / \sigma} \left( \sigma \sqrt{\omega} x \right) 
      , \quad \rho = +1 , \\*[9pt]
      \sinh^{-1 / \sigma} \left( \sigma \sqrt{\omega} x \right) 
      , \quad \rho = -1 ,
   \end{array} \right. 
\end{equation}
where $A_0 \equiv \left[ (1+\sigma) \omega \right]^{1/{2 \sigma }}$.
Solution~(\ref{eq:u0_power}) for $\rho=+1$ is the well-known bright
soliton of the generalized NLS equation that exists for any positive
value of $\sigma$. The solution for $\rho=-1$ is singular, since bright
solitons do not exist in a self-defocusing medium.
This singular solution will be used below to construct the profiles of
the (regular) localized modes in the general case with an impurity.

To find the soliton power in a self-focusing medium ($\rho=+1$) we 
substitute the corresponding solution (\ref{eq:u0_power}) into 
Eq.~(\ref{eq:P0}) and obtain the result:
\begin{equation} \label{eq:P0_sf}
   P_0 (\omega) =  \omega^{\frac{2-\sigma}{2\sigma}}   
   \frac{2^{2/\sigma}(1+\sigma)^{1/\sigma}}{2\sigma}
   \frac{\Gamma_S^2(1/\sigma)}{\Gamma_S(2/\sigma)} ,
\end{equation}
where $\Gamma_S$ is the standard Gamma-function.  Applying the VK theorem~\cite{vk} we find that the solitons are stable for $\sigma<\sigma_{\rm cr} \equiv 2$, because in this case $d P_0 / d \omega$ is positive. For $\sigma\ge2$ the solitons are unstable ($dP_0 / d \omega \le 0$) and they either collapse or decay~\cite{berge,pelin}.

\subsection{Nonlinear impurity in a linear medium}
            \label{sect:power_nl_imp}

We now study the effects due to defect-induced localization only, i.e. we consider the special case when a nonlinear defect with the power law-response
\begin{equation} \label{eq:NG_power_G}
   G(I) = \alpha + \beta I^{\gamma} , \quad \gamma>0,
\end{equation}
is embedded into a linear medium with $F \equiv 0$. Changing the signs of 
the parameters $\alpha$ and $\beta$ we can describe both linearly and 
nonlinearly attractive and/or repulsive impurities. Assuming that 
$\alpha\ne 0$ this parameter can be rescaled to $\alpha=\pm 2$, which 
we use in the following.

As bright solitons do not exist in a linear medium, asymmetric modes are
not possible as well.  Using Eqs.~(\ref{eq:u}),(\ref{eq:u_slv}), and~(\ref{eq:x0}) we obtain the spatial profile of the symmetric one-hump localized modes 
\begin{equation} \label{eq:prof_linear}
   u_i(x) =  A_i\exp(-\sqrt\omega |x|),  \quad  
   A_i^{2\gamma} = (2 \sqrt\omega - \alpha)/\beta,
\end{equation}
and the corresponding power,
\begin{equation} \label{eq:P_k_linear}
   P_{\rm i}(\omega) = \frac{A_i^2}{\sqrt\omega} 
   = \frac{1}{\sqrt\omega} {\left( \frac{2\sqrt\omega - \alpha}{\beta}  
     \right)}^{{1}/{\gamma}}.
\end{equation}
Let us analyze the existence properties of these modes:  
For $\alpha>0$ a {\em linear impurity mode} appears at the {\em cut-off 
frequency} $\omega_0=\alpha^2/4$.  If the nonlinearity is attractive 
($\beta>0$) then a whole family of localized modes exist with frequencies
above cut-off ($\omega\ge\omega_0$).  If the nonlinearity is repulsive 
($\beta<0$) then the family exists below cut-off ($0<\omega \le\omega_0$).
For $\alpha<0$ no localized modes exist for repulsive nonlinearities
($\beta<0$), whereas attractive nonlinearities ($\beta>0$) supports
localized modes at all frequencies $\omega>0$.

To analyse the stability properties we follow the approach of Sec.~\ref{sect:stab_symm} and calculate the defect response 
\begin{equation} \label{eq:G_linear}
   G_1 = G_1^{\rm cr} + 2 \beta \gamma A_i^{2\gamma},
\end{equation}
where the critical value $G_1^{\rm cr}=2\sqrt\omega$ is defined from Eq.~(\ref{eq:M1_cr}). From the general results summarized in Table~\ref{tab:stab_symm}, we see that the localized modes are {\em always stable} when the nonlinearity is repulsive ($\beta<0$), since $G_1\le G_1^{\rm cr}$. When the nonlinearity is attractive ($\beta>0$), $G_1 > G_1^{\rm cr}$ and the VK criterion applies. From Eq.~(\ref{eq:P_k_linear}) we obtain that the sign of $dP_i/d\omega$ is given by the sign of $[2(1-\gamma)\sqrt\omega +\alpha\gamma$]. Thus we identify the critical power $\gamma_{\rm cr}=1$, which is {\em half that in a homogeneous medium}. For $\alpha>0$ (and $\beta>0$) the modes are therefore always stable for sub-critical powers $\gamma<\gamma_{\rm cr}$. For $\gamma\ge \gamma_{\rm cr}$ high frequency modes with $\omega\ge\omega_1\equiv\omega_0 /(1-\gamma_{\rm cr}/\gamma)^2$ are unstable, whereas low frequency modes ($\omega<\omega_1$) are stable. The opposite occurs for $\alpha<0$ (and $\beta>0$): In this case only high frequency modes ($\omega>\omega_1$) are stable for $\gamma<\gamma_{\rm cr}$, whereas low frequency modes are unstable. For powers above the critical value, $\gamma\ge\gamma_{\rm cr}$, all modes are unstable.

\subsection{Nonlinear impurity modes} \label{sect:power_nl_full}

We now consider a more general case when both the bulk medium and the defect have power-law nonlinear responses, as defined by Eqs.~(\ref{eq:NG_power_F}) and (\ref{eq:NG_power_G}).

{\em Symmetric nonlinear modes:}
Substituting Eq.~(\ref{eq:u0_power}) into Eq.~(\ref{eq:x0}) we obtain the 
relation:
\begin{equation} \label{eq:x0_power}
   2p\omega^{1/2} + \alpha + \beta\omega^\Gamma (1 +\sigma)^\Gamma
   |1-p^2|^\Gamma = 0 , 
\end{equation}
which defines the spatial shift $x_0$ for all values of the mode frequency 
$\omega$.  Here $\Gamma=\gamma/\sigma$ and
\begin{equation} \label{eq:p}
   p(\omega) = \left\{ 
   \begin{array}{l}
      \tanh {\left( \sigma {\omega}^{1/2} x_0 \right)}, 
                                          \quad \rho = +1 ,\\*[5pt]
      \coth {\left( \sigma {\omega}^{1/2} x_0 \right)}, 
                                          \quad \rho = -1 . 
   \end{array} \right.
\end{equation}
We see that for $|p| < 1$, the solutions of Eq.~(\ref{eq:x0_power})
correspond to the modes in a {\em self-focusing} medium ($\rho = +1$). Note that localization in a {\em defocusing} medium ($\rho=-1$) is only supported by an attractive impurity ($G_0>0$), so that $x_0<0$ and $p<-1$. In both cases the linear limit $x_0 \rightarrow -\infty$ corresponds to $p \rightarrow -1$. Impurity response for a localized mode can be expressed in terms of the new variable, 
\begin{equation} \label{eq:G_pnl}
 \begin{array}{l}
 {\displaystyle
   G_0 = -2 p \sqrt{\omega}, 
 } \\*[9pt] {\displaystyle
   G_1 = G_0 + 2 \beta \gamma I_0^\gamma 
       = (2 \gamma+1) G_0 - 2 \gamma \alpha ,
 } \\*[9pt] {\displaystyle
   G_1^{\rm cr} = - 2 \sqrt{\omega} 
                    \left[ p (1+\sigma) - \sigma / p \right] ,
 } \end{array}
\end{equation}
which will be useful in the following analysis. 

In order to determine the stability of the localized modes, 
we have to construct the power vs. frequency diagram $P(\omega)$. 
After a change of variables in Eq.~(\ref{eq:Pm}) we obtain the general expression for the power
\begin{equation} \label{eq:P_p}
   P(\omega) = {\omega}^{\frac{2-\sigma}{2\sigma}} 
               \frac{2 (1+\sigma)^{1/\sigma}}{\sigma}
               \left| \int_{-p}^1 |1-y^2|^{-1+1/\sigma} dy \right|.
\end{equation}
To separate the effects induced by the bulk and the impurity we re-write this expression in the form:
\begin{equation} \label{eq:P_kp}
   P(\omega) = P_0 (\omega) \xi(p;\sigma) ,
\end{equation}
where the soliton power in homogeneous media, $P_0(\omega)$, is defined by Eq.~(\ref{eq:P0_sf}). Although bright solitons do not exist in {\em defocusing} homogeneous media, Eqs.~(\ref{eq:P0_sf}) and (\ref{eq:P_kp}) can still be used for both types of bulk nonlinearity, i.e.~for $\rho=\pm 1$. The functional $\xi(p, \sigma)$ changes monotonically with the parameter $p$, so that $\rho(\partial\xi/\partial p)>0$. In the linear limit when $\alpha>0$ and $\omega\rightarrow\omega_0$ the mode power vanishes and thus $\xi(-1;\sigma)=0$. Furthermore, for a self-focusing bulk medium ($\rho=+1$) the identity $\xi(-p;\sigma)=2-\xi(p;\sigma)$ holds. Thus $\xi(0;\sigma)=1$ and $\xi(1;\sigma)=2$. The functional $\xi(p;\sigma)$ cannot be expressed in elementary functions, but must be calculated numerically. 

Let us outline some general properties of the symmetric modes:
Consider first linearly attractive impurities ($\alpha>0$) and the 
{\em low-intensity limit} when the mode frequency is close to the 
cut-off $\omega_0$.
From Eq.~(\ref{eq:x0_power}) we find the {\em nonlinearity-induced 
frequency shift}
\begin{eqnarray} \label{eq:omega_shift}
   \left( \frac{\omega}{\omega_0} - 1 \right) \simeq \left\{ 
   \begin{array}{l}
       \beta 2^\Gamma (1+\sigma)^\Gamma \omega_0^{\Gamma-1/2}
       |1+p|^\Gamma , \,\, \Gamma < 1 ,\\
       2 (1+p) , \quad \Gamma > 1 .
   \end{array} 
   \right.
\end{eqnarray}
It follows that for $\Gamma>1$ the sign of the shift depends only on the bulk nonlinearity: the frequency is shifted up in self-focusing bulk media  ($\rho=+1$) and down in self-defocusing bulk media ($\rho=-1$). For $\Gamma<1$ the shift depends only on the defect nonlinearity: the frequency is shifted up when $\beta>0$ and down when $\beta<0$. However, Eq.~(\ref{eq:omega_shift}) presents only an asymptotic result valid at vanishingly small intensities. A simple graphical analysis of Eq.~(\ref{eq:x0_power}) demonstrates that competition of the defect and bulk nonlinearities for $\rho \beta < 0$ can lead to a more complicated (e.g. multi-valued) structure of the power dependence $P(\omega)$.

Secondly, we analyze the properties of {\em high-frequency modes} in 
self-focusing media ($\rho=+1$). Analysis of Eq.~(\ref{eq:x0_power}) reveals that such modes always exist. Substituting approximate solutions for $p(\omega)$ when $\omega\rightarrow \infty$ into Eq.~(\ref{eq:P_p}) we find the power of the high-frequency modes
\begin{equation} \label{eq:P_inf}
   P(\omega) \rightarrow \left\{ 
   \begin{array}{l}
      P_0(\omega) (1-\beta C\omega^{\Gamma-1/2} ) ,\quad \Gamma<1/2 ,\\
      P_{\rm i}(\omega) ,\quad \Gamma>1/2 \quad {\rm and} \quad \beta>0 , \\
      2P_0(\omega) ,\quad   \Gamma>1/2 \quad {\rm and} \quad \beta<0 ,
   \end{array} 
   \right.
\end{equation}
where $C$ is a frequency-independent positive coefficient. It follows from Eq.~(\ref{eq:P_inf}) that for a nonlinearly repulsive defect ($\beta<0$) we have $P(\omega)>P_0(\omega)$ and thus the high-frequency symmetric modes have two humps and are unstable. For nonlinearly attractive defects ($\beta>0$) the modes have one hump, i.e. $-1 < p < 0$. It is straightforward to check using Eq.~(\ref{eq:G_pnl}) that under these conditions $G_1 > G_0 > G_1^{\rm cr}$, and therefore according to Table~\ref{tab:stab_symm} the mode stability follows directly from the power slope. In particular, if $\Gamma<1/2$ then the nonlinearity of the bulk is effectively stronger than that of the defect, so that the localized modes resemble the solitons in homogeneous media, which are stable for $\sigma<2$. In the opposite case, $\Gamma>1/2$, the modes resemble the nonlinear defect modes of linear media, which are stable for $\gamma<1$. Combining these results we come to the conclusions: Symmetric localized modes supported by a nonlinearly attractive impurity ($\beta>0$) in a self-focusing bulk medium are stable in the high-frequency limit if $\sigma<2$ and $\gamma<1$ simultaneously.

Finally, we study the properties of high-frequency localized modes in self-defocusing bulk media ($\rho=-1$). In this case Eq.~(\ref{eq:x0_power}) has solutions for $\omega\rightarrow +\infty$ only for a ``strongly'' nonlinear attractive impurity with $\Gamma>1/2$ and $\beta>0$.  The corresponding modes have the same properties as in a self-focusing bulk medium, because the bulk nonlinearity acts only as a small perturbation. Indeed, according to Eq.~(\ref{eq:G_pnl}) we have $G_1 - G_1^{\rm cr} = - 2 \sqrt{\omega} [ \sigma/p + p (2 \gamma-\sigma) ] - 2 \gamma \alpha > 0$ for $\omega \gg 1$, and therefore the mode stability follows from the power slope.

{\em Asymmetric nonlinear modes} :
Asymmetric modes can exist in self-focusing bulk media when the impurity 
response vanishes, $G(I_0)=0$. This is possible when $\alpha\beta<0$ and the mode intensity at the defect site is $I_0=|\alpha/\beta|^{1/\gamma}$. Then, the spatial shift $x_0$ is given by the relation
\begin{equation}
   \cosh(\sigma\sqrt\omega x_0) = A_0^\sigma/I_0^{\sigma/2} .
\end{equation}
Asymmetric modes bifurcate from the symmetric modes at the bifurcation point $\omega_b=I_0^{\sigma}/(1+\sigma)$ when $A_0^2=I_0$ and they exist for higher frequencies only. An asymmetric mode is actually a soliton trapped by the defect and is therefore only stable for $\sigma<2$. However, the trapping (defined by the condition $G_0(I_0)=0$) is stable if $G_1 = 2 \beta \gamma I_0 < 0$ (see Sec.~\ref{sect:stab_asymm}), which is satisfied for nonlinearly repulsive impurities with $\beta<0$.

\section{Self-focusing medium}  \label{sect:pw_foc}

In this section we present a detailed numerical analysis of the nonlinear impurity modes for power-law nonlinearities in a {\em self-focusing bulk medium} ($\rho=+1 $). The self-defocusing bulk medium is treated in Sec.~\ref{sect:pw_defoc}. We term the impurity {\em attractive} when it supports a localized mode in the linear limit ($\alpha>0$) and has a self-focusing (attractive) nonlinear response ($\beta>0$). This case is considered in Sec.~\ref{sect:pw_foc_ai}. The same impurity, but with a self-defocusing nonlinear response ($\beta<0$) is termed a {\em mixed impurity that supports linear modes} (see Sec.~\ref{sect:pw_foc_mi}). The case when the defect is repulsive in the linear limit ($\alpha<0$), but may become attractive for larger intensities ($\beta>0$) due to its nonlinear properties is the most complicated. This {\em mixed impurity with attractive nonlinearity} is investigated in Sec.~\ref{sect:pw_foc_ri}. When both $\alpha$ and $\beta$ are {\em negative} the properties of localized modes in self-focusing bulk media are trivial: they always have two-hump symmetric profiles and are therefore always unstable. 

\subsection{Attractive impurity} \label{sect:pw_foc_ai}

\begin{figure}
   \setlength{\epsfxsize}{8.0cm}
   \vspace*{-20mm}
   \centerline{\mbox{\epsffile{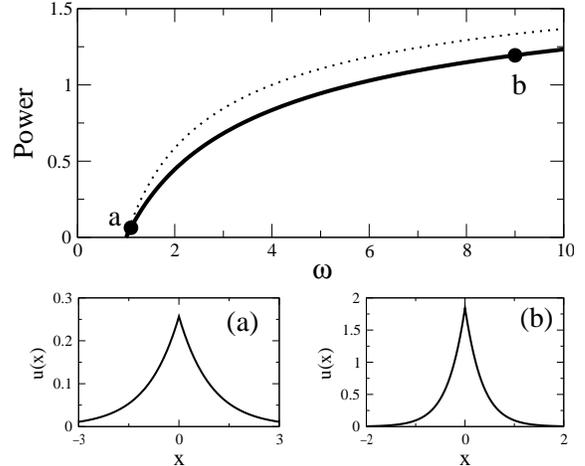}}}
   \vspace*{-20mm}
   \caption{ \label{fig:pwr-sf-al_p-beta_p_s1g1}
   Power vs.~frequency diagram for $\gamma/\sigma = 1$ and two characteristic 
   mode profiles corresponding to the points (a) and (b).
   Dotted curve: power $P_{\rm i}(\omega)$ of impurity modes in linear 
   bulk media. Parameters: $\sigma = 1$, $\gamma = 1$, $\alpha = 2$, $\beta = 1$.} 
\end{figure}

\begin{figure}
  \setlength{\epsfxsize}{8.0cm}
  \vspace*{-25mm}
  \centerline{\mbox{\epsffile{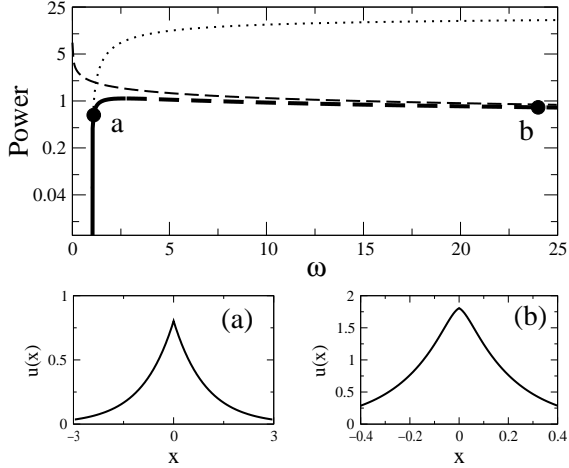}}}
  \vspace*{-25mm}
  \caption{\label{fig:pwr-sf-al_p-beta_p_s4g1}
  Power vs.~frequency diagram for $\gamma/\sigma = 1/4$. Solid and dashed 
  thick curves show stable and unstable modes, respectively. 
  Thin dashed curve: soliton power $P_0(\omega)$. 
  Thin dotted curve: power $P_{\rm i}(\omega)$ of impurity modes in linear 
  bulk media.  Modes corresponding to points (a) and (b) are shown below.
  Parameters: $\sigma = 4$, $\gamma = 1$, $\alpha = 2$, $\beta = 0.1$.} 
\end{figure}

Let us consider self-focusing bulk media ($\rho=+1$) and defects that 
support localized modes in the linear limit ($\alpha>0$). 
In this case the symmetric mode branch on the $P(\omega)$ diagram always 
starts at $\omega=\omega_0$, where $P$ vanishes.
As follows from the stability analysis this point is critical, and 
therefore the localized modes are always stable in its close vicinity, 
see Figs.~\ref{fig:pwr-sf-al_p-beta_p_s1g1} and~\ref{fig:pwr-sf-al_p-beta_p_s4g1}.
However, for higher frequencies $\omega$ the properties of the localized modes depend on the characteristics of the nonlinearity, determined by the powers $\gamma$ and $\sigma$ and the coefficient $\beta$.

\begin{figure}
  \setlength{\epsfxsize}{7.0cm}
  \vspace*{-25mm}
  \centerline{\mbox{\epsffile{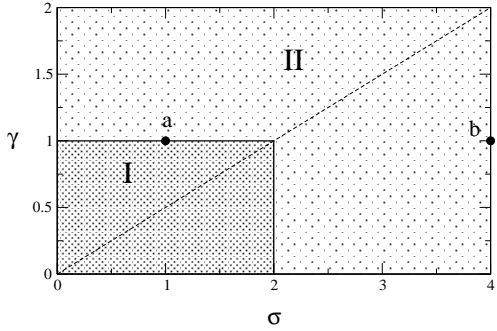}}}
  \vspace*{-25mm}
  \caption{ \label{fig:stab-sf-al_p-beta_p}
  Stability regions for symmetric localized modes with $\alpha>0$ and 
  $\beta>0$ in self-focusing media ($\rho=+1$): 
  I:~the modes are stable and exist for $\omega>\omega_0$; 
  II:~stable modes exist only near 
  cut-off $\omega\simeq\omega_0$.  Points~(a) and~(b) correspond 
  to the cases shown in Figs.~\ref{fig:pwr-sf-al_p-beta_p_s1g1} and 
  \ref{fig:pwr-sf-al_p-beta_p_s4g1}, respectively.}
\end{figure}

We first consider the influence of defect nonlinearities that enhance the attraction ($\beta>0$). In this case the asymmetric modes do not exist. For the symmetric localized states we always have $x_0<0$ and, as follows from Eq.~(\ref{eq:G_pnl}), $G_1 > G_0 > G_1^{\rm cr}$, so that the VK criterion \cite{vk} applies according to Table~\ref{tab:stab_symm}. To illustrate the general results given in Sec.~\ref{sect:power_nl_full}, we present examples in Figs.~\ref{fig:pwr-sf-al_p-beta_p_s1g1} and~\ref{fig:pwr-sf-al_p-beta_p_s4g1}, and the stability diagram in Fig.~\ref{fig:stab-sf-al_p-beta_p}. Localized modes always exist above cut-off, $\omega\ge\omega_0$, and the functional $P(\omega)$ is single-valued. Such a behavior is observed because both the defect and the bulk medium have self-focusing nonlinearities, which induce a positive frequency shift, as mentioned earlier.

The parameters for Figs.~\ref{fig:pwr-sf-al_p-beta_p_s1g1} and~\ref{fig:pwr-sf-al_p-beta_p_s4g1} correspond to {\em qualitatively different cases}. For $\gamma/\sigma>1/2$ (Fig.~\ref{fig:pwr-sf-al_p-beta_p_s1g1}) the defect nonlinearity is effectively stronger than the bulk nonlinearity, and thus the modes resemble the nonlinear defect modes in linear bulk media. Since $\gamma$ and $\sigma$ do not exceed their critical values $\gamma_{\rm cr} = 1$ and $\sigma_{\rm cr} = 2$ [see Fig.~\ref{fig:stab-sf-al_p-beta_p}, point (a)], all modes should be stable. For $\gamma/\sigma<1/2$, the high-frequency symmetric modes resemble the solitons of homogeneous media [see Fig.~\ref{fig:pwr-sf-al_p-beta_p_s4g1}(b)]. Therefore, these modes are  unstable if the bulk nonlinearity is super-critical $\sigma>\sigma_{\rm cr}$ [see Fig.~\ref{fig:stab-sf-al_p-beta_p}].

\begin{figure}
  \setlength{\epsfxsize}{8.0cm}
  \vspace*{-1cm}
  \centerline{\mbox{\epsffile{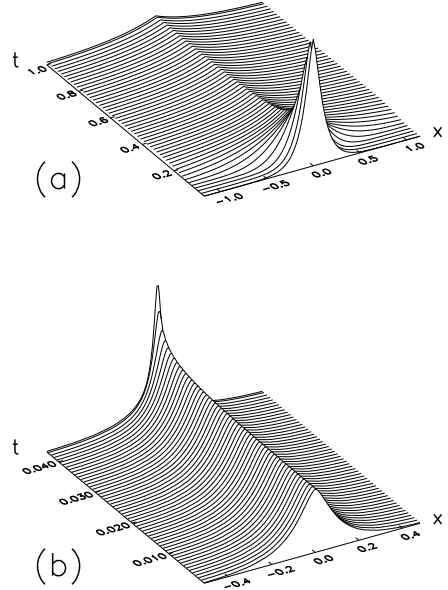}}}
  \vspace{2mm}
  \caption{ \label{fig:bpm-sf-al_p-beta_p}
  Evolution of an unstable mode [Fig.~\ref{fig:pwr-sf-al_p-beta_p_s4g1}(b)]   with the input power decreased (a) or increased (b) by 0.1\%. 
  Two different scenarios of the instability-induced mode dynamics are
  observed:
  (a)~switching from an unstable to a stable state, and
  (b)~collapse induced by the bulk nonlinearity. }
\end{figure}

To study the instability-induced dynamics of symmetric localized modes we solve Eq.~(\ref{eq:nls}) numerically with slightly perturbed modes as initial condition. Depending on the perturbation, {\em two different scenarios} are possible. If the power is initially decreased, then the mode spreads out and transforms into a lower-frequency stable mode, as shown in Fig.~\ref{fig:bpm-sf-al_p-beta_p}(a). We note that this switching process is accompanied by some power loss due to radiation. An initial increase of the mode power can lead to collapse if the nonlinear self-focusing dominates the linear diffraction. Figure \ref{fig:bpm-sf-al_p-beta_p}(b) shows a collapsing mode, whose amplitude goes to infinity in a finite time due to the effect of a supercritical bulk nonlinearity, $\sigma\ge2$, even though the collapse occurs at the impurity site. This collapse instability was earlier investigated for uniform nonlinear media \cite{berge}, with many of the general features applying to the impurity modes as well. For example, the collapsing solution typically consists of a slowly evolving background and a highly-localized central part having an almost self-similar profile. 

\subsection{Mixed impurity that supports linear modes}
            \label{sect:pw_foc_mi}

\begin{figure}
  \setlength{\epsfxsize}{8.0cm}
  \vspace*{-25mm}
  \centerline{\mbox{\epsffile{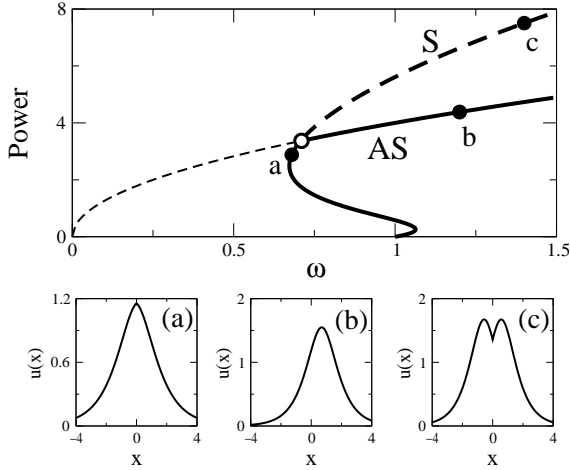}}}
  \vspace*{-25mm}
  \caption{ \label{fig:pwr-sf-al_p-beta_m_s1g2} 
  Power vs.~frequency diagram for $\gamma > \sigma$ ($\sigma = 1$, 
  $\gamma = 2$, $\alpha = 2$, $\beta = -1$, $\beta_{\rm cr} \simeq -0.21$).
  Solid and dashed thick curves show stable and unstable localized modes, 
  respectively.
  Open circle: bifurcation point from symmetric (S) to asymmetric (AS) modes.
  Thin dashed curve: soliton power $P_0(\omega)$. 
  The modes corresponding to points (a), (b), and (c) are shown below.}
\end{figure}

\begin{figure}
  \setlength{\epsfxsize}{8.0cm}
  \vspace*{-25mm}
  \centerline{\mbox{\epsffile{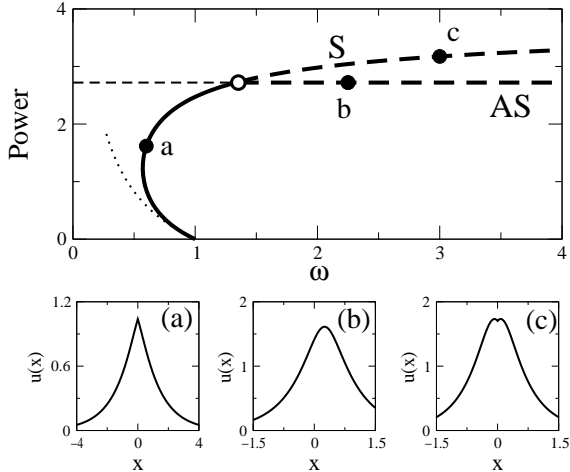}}}
  \vspace*{-25mm}
  \caption{ \label{fig:pwr-sf-al_p-beta_m_s2g1} 
  Power vs.~frequency diagram for $\gamma<\sigma$ ($\sigma = 2$, 
  $\gamma = 1$).
  Parameters and notations are the same as in 
  Fig.~\ref{fig:pwr-sf-al_p-beta_m_s1g2}.
  Dotted curve: impurity mode power in a linear medium, $P_{\rm i}(\omega)$.}
\end{figure}

We now consider defects that have a {\em repulsive nonlinearity}, $\beta<0$ (and still $\alpha>0$). Due to the competition between focusing bulk and defocusing defect nonlinearities the power functional of the {\em symmetric} modes can become {\em multi-valued}, with two or three states having the same frequency (see Figs.~\ref{fig:pwr-sf-al_p-beta_m_s1g2} and \ref{fig:pwr-sf-al_p-beta_m_s2g1}). Such states can even exist below the linear cut-off frequency, i.e.~for $\omega<\omega_0$. This is always possible for $\gamma<\sigma$, because the branch starting at the critical point $P=0$ has a negative slope (still stable since $G_1<G_1^{\rm cr}$), and then the slope changes sign as the branch goes through another critical point, as shown in Fig.~\ref{fig:pwr-sf-al_p-beta_m_s2g1}. For $\gamma>\sigma$ the initial slope is positive, but then the branch can go through two critical points, which appear if $\beta<\beta_{\rm cr}$ (see Fig.~\ref{fig:pwr-sf-al_p-beta_m_s1g2}), where 
\begin{equation} \label{eq:beta_cr}
   \beta_{\rm cr} = - |\alpha|^{1-2\Gamma}
         \frac{(\Gamma-1)}{(2 \Gamma - 1)}
         {\left| \frac{2(2 \Gamma-1)^2}{(1+\sigma)\Gamma^2(\Gamma-1)} 
	 \right|}^{\Gamma} .
\end{equation}

The branch of asymmetric modes emerges at the bifurcation point where 
$P(\omega)$ coincides with the power of solitons in homogeneous media, 
$P_0(\omega)$ (open circles in Figs.~\ref{fig:pwr-sf-al_p-beta_m_s1g2} 
and \ref{fig:pwr-sf-al_p-beta_m_s2g1}). 
As demonstrated in Sec.~\ref{sect:power_nl_full} the stability of 
asymmetric modes for $\beta<0$ is the same as that of solitons 
in bulk media (see Sec.~\ref{sect:power_nl_sol}), i.e.~the modes 
are stable for $\sigma<2$ and unstable otherwise. 
After the bifurcation point, i.e.~for $\omega>\omega_b$, the symmetric 
localized mode becomes two-humped and therefore unstable.

\begin{figure}
  \setlength{\epsfxsize}{8.0cm}
  \vspace*{-10mm}
  \centerline{\mbox{\epsffile{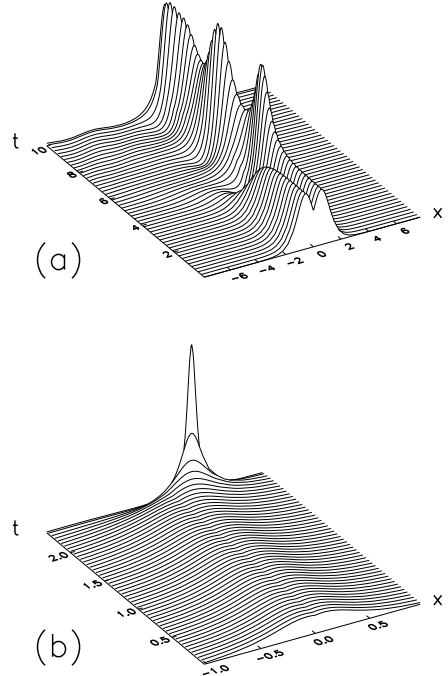}}}
  \vspace{2mm}
  \caption{ \label{fig:bpm-sf-al_p-beta_m}
  (a,b) Evolution of perturbed unstable two-hump symmetric localized 
  modes shown in Figs.~\ref{fig:pwr-sf-al_p-beta_m_s1g2}(c) and 
  \ref{fig:pwr-sf-al_p-beta_m_s2g1}(c), respectively.
  In (a) the mode transforms into a stable asymmetric mode;
  in (b) the transition occurs to an unstable asymmetric mode, which 
  subsequently collapses in the bulk medium. }
\end{figure}

In Figs.~\ref{fig:bpm-sf-al_p-beta_m}(a) and (b) we show the evolution of 
two-hump symmetric modes when the bulk nonlinearity is sub- and super-critical,
respectively.  In both cases a symmetry breaking occurs, and the mode is 
repelled to one side of the defect. 
In case (a) a stable asymmetric state is excited (as $\sigma<2$), together 
with its internal mode, leading to a slowly decaying quasi-periodic beating. 
On the other hand, if bright solitons are unstable, the mode collapses in 
the bulk medium as shown in Fig.~\ref{fig:bpm-sf-al_p-beta_m}(b). 
Note, that the amplitude at the impurity always remains finite due to the 
repulsive nonlinear response of the defect. 

\subsection{Mixed impurity with attractive nonlinearity} 
            \label{sect:pw_foc_ri}

\begin{figure}
  \setlength{\epsfxsize}{8.0cm}
  \vspace*{-25mm}
  \centerline{\mbox{\epsffile{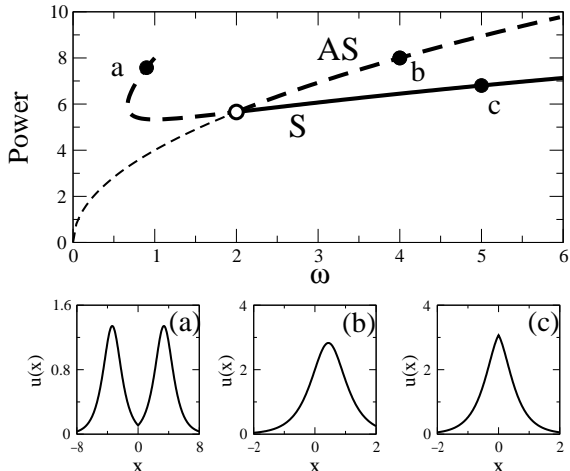}}}
  \vspace*{-25mm}
  \caption{ \label{fig:pwr-sf-al_m-beta_p_s1g05}
  Power vs.~frequency diagram for $\sigma = 1$, $\gamma = 0.5$, 
  $\beta = 1$, and $\alpha = -2$. Notations are the same as in 
  Fig.~\ref{fig:pwr-sf-al_p-beta_m_s1g2}.}
\end{figure}

\begin{figure}
  \setlength{\epsfxsize}{8.0cm}
  \vspace*{-25mm}
  \centerline{\mbox{\epsffile{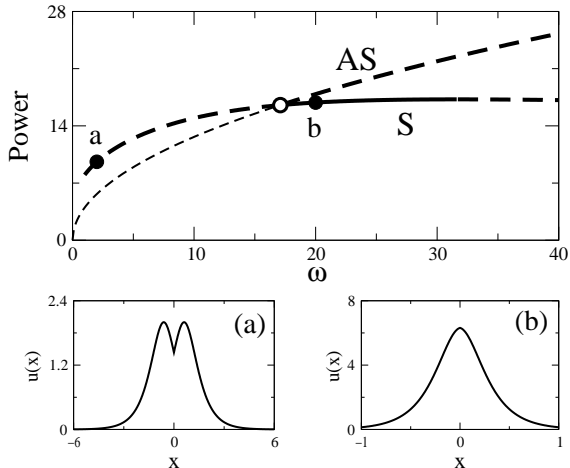}}}
  \vspace*{-25mm}
  \caption{ \label{fig:pwr-sf-al_m-beta_p_s1g15}
  Power vs.~frequency diagram for $\sigma = 1$, $\gamma = 1.5$,
  $\beta = 0.01$, and $\beta_{\rm max} \simeq 0.0262$.  
  Parameters and notations are the same as in 
  Fig.~\ref{fig:pwr-sf-al_m-beta_p_s1g05}. }
\end{figure}

\begin{figure}
  \setlength{\epsfxsize}{8.0cm}
  \vspace*{-25mm}
  \centerline{\mbox{\epsffile{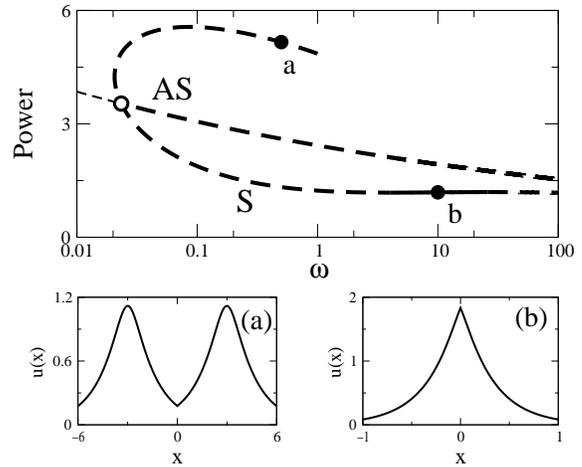}}}
  \vspace*{-25mm}
  \caption{ \label{fig:pwr-sf-al_m-beta_p_s25g05}
  Power vs.~frequency diagram for $\sigma = 2.5$, $\gamma = 0.5$, 
  $\beta = 3.3$, and $\beta_{\rm min} \simeq 3.17$.  Parameters and notations are the same as in 
  Fig.~\ref{fig:pwr-sf-al_m-beta_p_s1g05}. }
\end{figure}

We finally consider the most complicated case of a linearly repulsive, but nonlinearly attractive defect ($\alpha<0$, $\beta>0$). Some properties of the {\em symmetric localized modes} are revealed by the symmetry constraints following Eqs.~(\ref{eq:p}) and (\ref{eq:P_kp}). Specifically, under an inversion of the defect response, $(\alpha,\beta) \rightarrow (-\alpha,-\beta)$, the normalized shift $p(x_0)$ and power $P$ change as follows: $p\rightarrow (1-p)$ and $P(\omega)\rightarrow 2P_0(\omega)-P(\omega)$. Applying this transformation to the case $\alpha<0$ and $\beta>0$ and using the results obtained above, we find that the branch of symmetric localized modes in the $P(\omega)$ functional starts at the point $P(\omega_0)=2P_0$. In the vicinity of this point the symmetric modes are unstable because $x_0>0$ (see Figs.~\ref{fig:pwr-sf-al_m-beta_p_s1g05}, \ref{fig:pwr-sf-al_m-beta_p_s1g15}, and~\ref{fig:pwr-sf-al_m-beta_p_s25g05}). At the bifurcation point (open circle) the family of asymmetric modes emerges, having the same power as solitons in a bulk medium, $P_0(\omega)$. The symmetric modes have one-hump profiles after the bifurcation. 

\begin{figure}
  \setlength{\epsfxsize}{7.0cm}
  \vspace*{-25mm}
  \centerline{\mbox{\epsffile{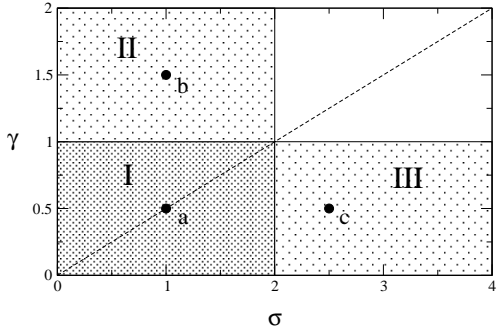}}}
  \vspace*{-25mm}
  \caption{ \label{fig:stab-sf-al_m-beta_p}
  Stability regions of localized modes for impurities with $\alpha>0$ and 
  $\beta>0$ in self-focusing bulk media:  I: the modes are stable for 
  frequencies above a threshold, $\omega>\omega_{\rm th}$; II and III: 
  stable localized states can exist in a bounded frequency range. 
  In the region without shading all modes are unstable.
  Parameter values at points (a), (b), and (c) correspond to 
  Figs.~\ref{fig:pwr-sf-al_m-beta_p_s1g05}, \ref{fig:pwr-sf-al_m-beta_p_s1g15},
  and \ref{fig:pwr-sf-al_m-beta_p_s25g05}, respectively.}
\end{figure}

The stability of symmetric one-hump localized modes depends on the nonlinearity characteristics. Since we have $\beta>0$ and $x_0<0$, it follows from Eq.~(\ref{eq:G_pnl}) that $G_1 > G_0 > G_1^{\rm cr}$, and stability is determined by the power slope according to Table~\ref{tab:stab_symm}. In particular, as demonstrated in Sec.~\ref{sect:power_nl_full}, the modes are stable in a range of frequencies unlimited from above if $\sigma<2$ and $\gamma<1$ (see, e.g., Fig.~\ref{fig:pwr-sf-al_m-beta_p_s1g05}). This parameter region~I in Fig.~\ref{fig:stab-sf-al_m-beta_p} is identical to that for $\alpha>0$ (see Fig.~\ref{fig:stab-sf-al_p-beta_p}); the mode properties are also similar, because in the limit $\omega \gg \omega_0$ the linear response of the defect acts as only a small perturbation. However, there is an important difference: while for $\alpha>0$ the modes are always stable for $\omega\simeq\omega_0$, this is not so for a linearly repulsive impurity. Although in the latter case the high-frequency modes are also unstable if the nonlinearity powers exceed critical values, we found that for the nonlinearity parameters corresponding to regions II or III in Fig.~\ref{fig:stab-sf-al_m-beta_p} the modes can still be stabilized at some frequencies.

The first mechanism of stabilization in a bounded frequency region above 
the bifurcation frequency $\omega_b$ is realized when the modes are close to stable solitons (for $\sigma<2$) [see Fig.~\ref{fig:pwr-sf-al_m-beta_p_s1g15}], and the strength of the defect is relatively weak (region~II in Fig.~\ref{fig:stab-sf-al_m-beta_p}). This is possible if the nonlinearity coefficient does not exceed a threshold value, which can be calculated from the condition $( d P / d \omega )_{\omega = \omega_b} > 0$,
\begin{equation}
  0 < \beta < \beta_{\rm max} 
      = \frac{ |\alpha| }{ (1+\sigma)^\Gamma }
     {\left[\frac{(2-\sigma) 2^{2/\sigma} \Gamma_S^2(1/\sigma)}{4\gamma 
     |\alpha| \Gamma_S(2/\sigma) } \right]}^{2 \Gamma} .
\end{equation}

\begin{figure}
  \setlength{\epsfxsize}{8.0cm}
  \vspace*{-25mm}
  \centerline{\mbox{\epsffile{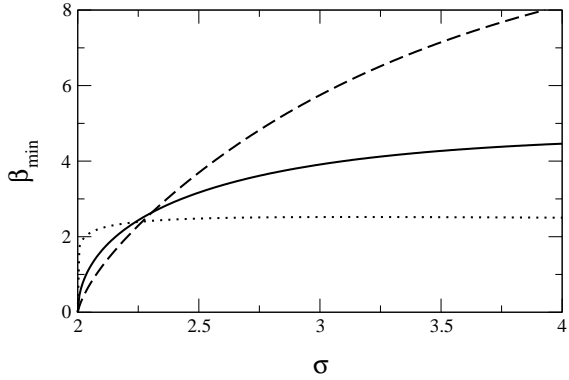}}}
  \vspace*{-25mm}
  \caption{ \label{fig:beta_min}
  Parameter $\beta_{\rm min}$ versus bulk power nonlinearity $\sigma$ for 
  $\gamma = 0.1$ (dotted), $0.5$ (solid), and $0.8$ (dashed).}
\end{figure}

On the other hand, if $\sigma>2$ the modes can be stabilized by the 
attractive defect nonlinearity, as demonstrated in Fig.~\ref{fig:pwr-sf-al_m-beta_p_s25g05}, provided the impurity supports non-collapsing highly localized states (i.e.~for $\gamma<1$, see region III in Fig.~\ref{fig:stab-sf-al_m-beta_p}), and has a sufficiently strong nonlinearity, $\beta>\beta_{\rm min}$. The minimum value $\beta_{\rm min}$ cannot be determined analytically. Its characteristic dependencies on $\sigma$ for several values of $\gamma$ are shown in Fig.~\ref{fig:beta_min}.

\begin{figure}
  \setlength{\epsfxsize}{8.0cm}
  \vspace*{-10mm}
  \centerline{\mbox{\epsffile{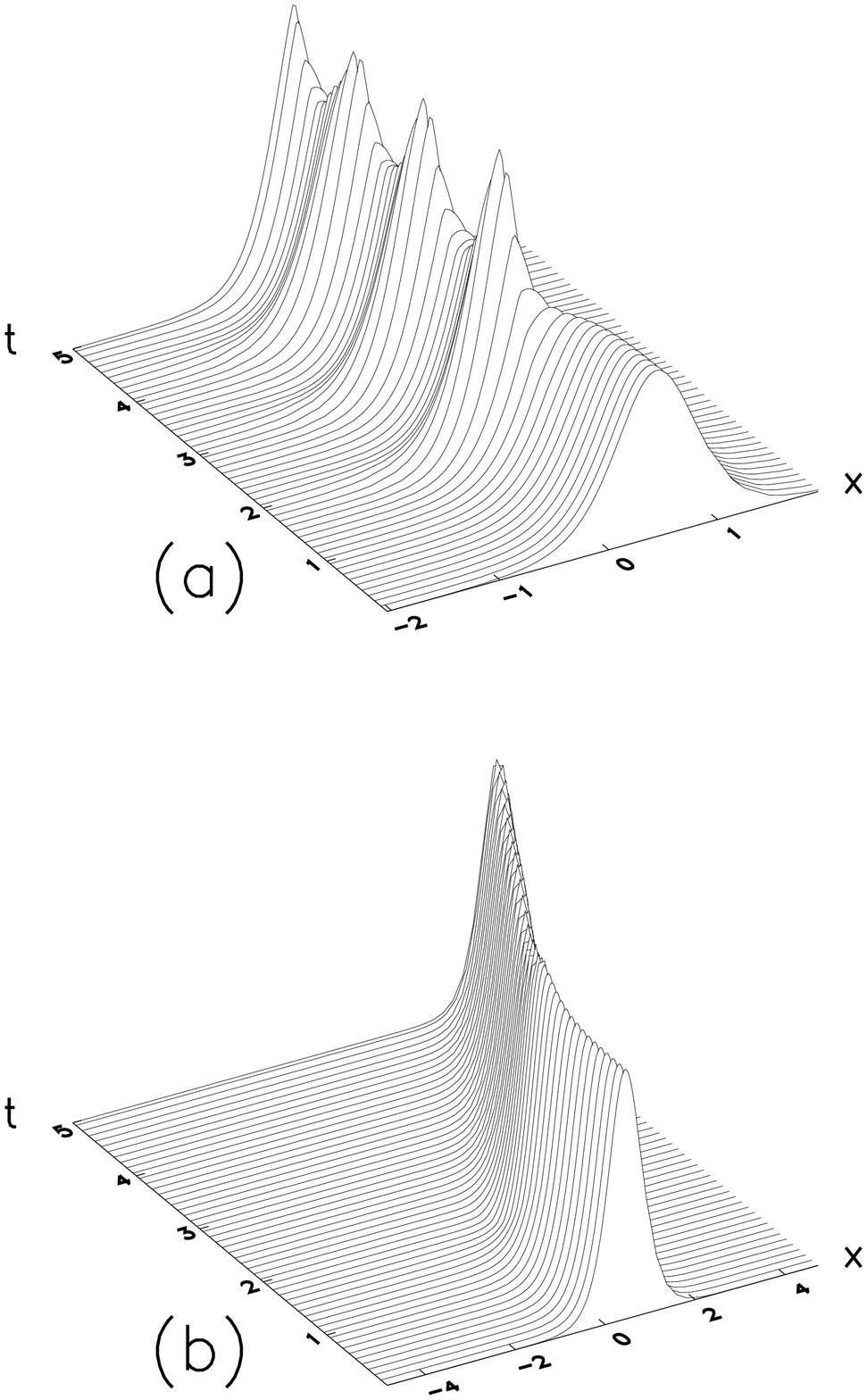}}}
  \vspace{2mm}
  \caption{ \label{fig:bpm-sf-al_m-beta_p-as}
  (a,b) Evolution of the perturbed asymmetric mode shown in 
  Fig.~\ref{fig:pwr-sf-al_m-beta_p_s1g05}(b). 
  (a) Attraction by the impurity and transformation into a stable 
  symmetric mode.
  (b) Repulsion by the impurity and transformation into a moving 
  soliton.}
\end{figure}

As demonstrated in Sec.~\ref{sect:power_nl_full} the asymmetric modes 
are unstable for $\beta>0$ with respect to a translational shift along the $x$ axis. To study the development of this instability we perform numerical simulations. The results confirmed that the mode is either attracted or repelled by the impurity, depending on the type of perturbation, as illustrated in Figs.~\ref{fig:bpm-sf-al_m-beta_p-as}(a) and~\ref{fig:bpm-sf-al_m-beta_p-as}(b), respectively. These examples correspond to the case of sub-critical defect and bulk nonlinearities, and thus the mode eventually transforms into a stable impurity mode or a moving soliton. However, for stronger nonlinearities the instability induced dynamics can result in a collapse.

\section{Self-defocusing medium}   \label{sect:pw_defoc}

In this section we focus on {\em self-defocusing} bulk media ($\rho=-1$). In contrast to self-focusing bulk media investigated above, bright solitons cannot exist in homogeneous self-defocusing bulk media, and thus localized modes appear solely due to the presence of the impurity, having {\em always a one-hump profile}.  Again we consider the different cases separately.

\subsection{Attractive impurity} \label{sect:pw_defoc_ai}

First we consider attractive defects with both $\alpha>0$ and $\beta>0$. 
It follows from the analysis presented above that localized modes exist 
and are stable near the cut-off frequency, $\omega\simeq\omega_0$. 
For defects with a purely linear response ($\gamma\rightarrow 0$) there 
exists a single branch of localized solutions on the $P(\omega)$ diagram 
in the region $0<\omega\le\omega_0$. This case was considered in 
Ref.~\cite{ltp} for $\sigma=1$.

\begin{figure}
  \vspace*{0mm}
  \setlength{\epsfxsize}{8.0cm}
  \centerline{\mbox{\epsffile{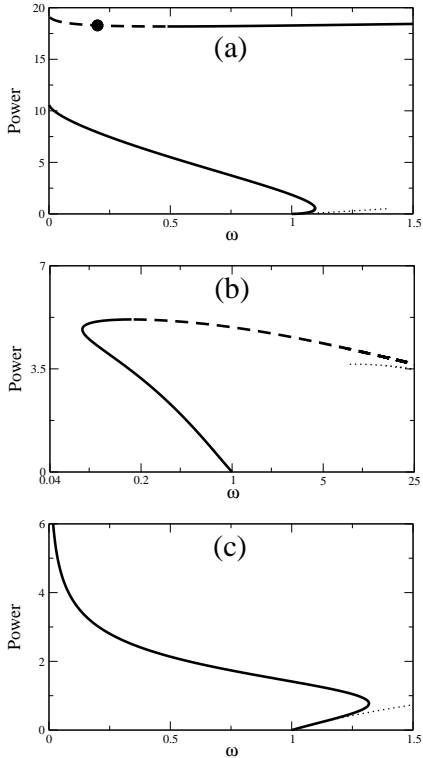}}}
  \vspace*{2mm}
  \caption{ \label{fig:pwr-df-al_p-beta_p}
  Power vs.~frequency diagrams for (a) $1/2<\Gamma< 1$ ($\sigma = 1$, 
  $\gamma = 0.7$, $\beta = 0.52$, $\beta_{\rm cr} \simeq 0.603$, and 
  $\beta_0 \simeq 0.533$);  (b) $\Gamma > 1$ ($\sigma = 1$, $\gamma = 1.5$,   $\beta = 0.11$, and $\beta_0 \simeq 0.105$); (c) $\Gamma<1/2$   
  ($\sigma = 3$, $\gamma = 1$, $\beta = 0.5$).
  Notations are the same as in Fig.~\ref{fig:pwr-sf-al_p-beta_p_s4g1}
  and $\alpha=2$.
  The marked point in (a) corresponds to an unstable solution whose
  instability dynamics is illustrated in Fig.~\ref{fig:stab-df-al_p-beta_p}.}
\end{figure}

Remarkably, defects with a nonlinear response can support localized modes 
with frequencies {\em above the linear cut-off frequency}.  Indeed, as we 
demonstrated in Sec.~\ref{sect:power_nl_full}, the initial slope of the 
$P(\omega)$ functional for weakly nonlinear modes is determined by the
nonlinearity of the impurity for $\Gamma<1$. 
If $\Gamma<1/2$ the branch goes through a critical point and disappears 
as $\omega\rightarrow 0$ [see Fig.~\ref{fig:pwr-df-al_p-beta_p}(c)], while the power $P$ remains bounded for $\sigma<2$, and is unbounded for $\sigma>2$. 
For $1/2<\Gamma<1$ the critical point appears only for $\beta<
\beta_{\rm cr}$, where $\beta_{\rm cr}$ is defined by Eq.~(\ref{eq:beta_cr}) 
[see Fig.~\ref{fig:pwr-df-al_p-beta_p}(a)]. 

\begin{figure}
  \setlength{\epsfxsize}{7.0cm}
  \vspace*{-25mm}
  \centerline{\mbox{\epsffile{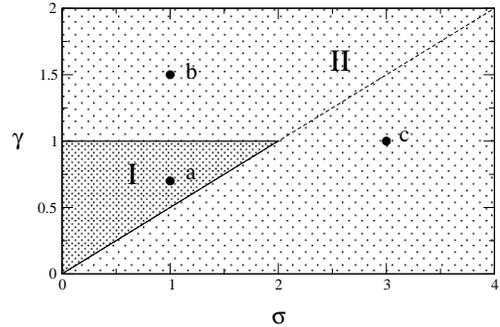}}}
  \vspace*{-25mm}
  \caption{ \label{fig:stab-df-al_p-beta_p}
  Stability regions of localized modes for defects with $\alpha>0$ and 
  $\beta>0$ in defocusing media.
  I: modes exist and are stable at both low and high frequencies;
  II: stable modes exists only close to and below the linear cut-off 
  $\omega\simeq\omega_0$. Points (a)-(c) correspond to 
  Figs.~\ref{fig:pwr-df-al_p-beta_p}(a)-(c), respectively.}
\end{figure}

If the nonlinear response of the impurity is effectively stronger than that 
of the bulk, highly localized high frequency modes can exist.  As was found 
in Sec.~\ref{sect:power_nl_full} this occurs for $\Gamma>1/2$. 
These modes have the power $P\simeq P_{\rm i}(\omega)$ when $\omega\gg
\omega_0$, and are thus stable for $\gamma<1$ (region I in 
Fig.~\ref{fig:stab-df-al_p-beta_p}). 
The mode frequency is bounded from below, $\omega>\omega_{\rm min}$, for 
$\beta>\beta_0$, where
\begin{equation} \label{eq:beta_0}
   \beta_0(\alpha,\sigma,\Gamma) = \frac{|\alpha|}{|2 \Gamma - 1| 
   (1+\sigma)^\Gamma} {\left| \frac{2 \Gamma - 1}{ \alpha \Gamma} 
   \right|}^{2 \Gamma},
\end{equation}
as shown in Fig.~\ref{fig:pwr-df-al_p-beta_p}(b). For $\beta<\beta_0$ the functional $P(\omega)$ has two branches originating at $\omega = 0$, with the upper branch corresponding to the highly localized modes discussed above. The lower branch approaches the linear limit $P(\omega_0) = 0$ and has initially a negative slope, which changes at the critical point for $1/2<\Gamma<1$, as shown in Fig.~\ref{fig:pwr-df-al_p-beta_p}(a) (consistent with the inequality $\beta_0<\beta_{\rm cr}$).

\begin{figure}
  \setlength{\epsfxsize}{8.0cm}
  \vspace*{-10mm}
  \centerline{\mbox{\epsffile{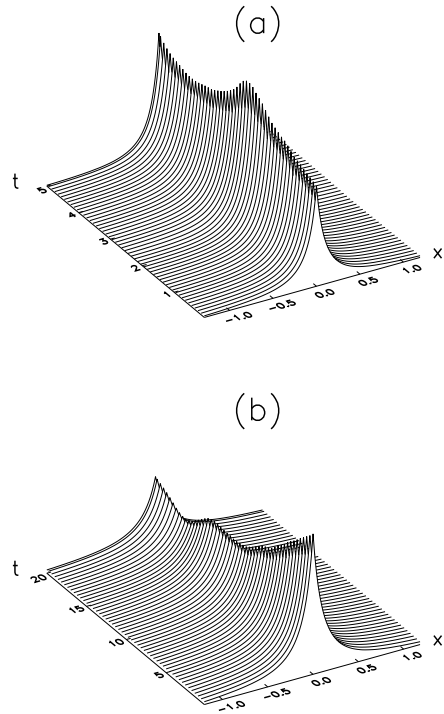}}}
  \vspace{2mm}
  \caption{ \label{fig:bpm-df-al_p-beta_p}
  Evolution of a perturbed asymmetric impurity mode corresponding to 
  the marked point in Fig.~\ref{fig:pwr-df-al_p-beta_p}(a).
  (a) Switching to a higher-frequency mode of the same branch.
  (b) A reduction of the power by 5\% leads to transformation to a 
  broader low-frequency mode of a lower branch.}
\end{figure}

To clarify some features of unstable modes in this case, we perform numerical simulations for the mode corresponding to the marked point in Fig.~\ref{fig:pwr-df-al_p-beta_p}(a). We find that all small perturbations result in switching to a more localized state of the same (upper) branch, as illustrated in Fig.~\ref{fig:bpm-df-al_p-beta_p}(a). However, if the power is decreased below the minimum of the upper branch, the mode evolves towards a stationary mode of the lowest branch [see Fig.~\ref{fig:bpm-df-al_p-beta_p}(b)]. In this case a substantial amount of power is radiated away due to the gap between the two branches.

\subsection{Mixed impurity that supports linear modes} 
            \label{sect:pw_defoc_mi}

\begin{figure}
  \vspace*{0mm}
  \setlength{\epsfxsize}{7.0cm}
  \vspace*{-25mm}
  \centerline{\mbox{\epsffile{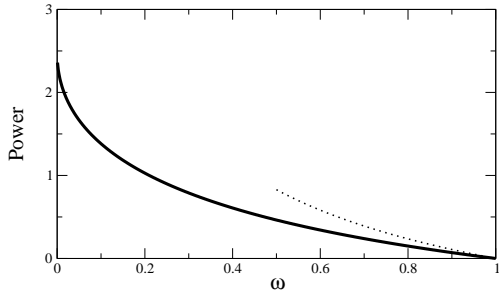}}}
  \vspace*{-25mm}
  \caption{ \label{fig:pwr-df-al_p-beta_m}
  Power vs.~frequency diagram for $\sigma = 1$, $\gamma = 1$, 
  $\alpha = 2$, and $\beta = -1$.
  Dotted curve: asymptotic given by the power $P_{\rm i}(\omega)$ of 
  linear impurity modes.}
\end{figure}

When both the bulk and defect have a self-defocusing nonlinear response 
($\rho=-1$, $\beta<0$) low-intensity localized modes can still exist when 
the linear attraction ($\alpha>0$) dominates the nonlinear de-localization 
effect. 
In this case the nonlinearity-induced frequency shift is {\em negative} 
(see Sec.~\ref{sect:power_nl_full}) and the localized solutions exist for 
frequencies below the linear cut-off, $\omega<\omega_0$. 
The corresponding dependence $P(\omega)$ is single-valued, as demonstrated 
in Fig.~\ref{fig:pwr-df-al_p-beta_m}. In the limit $\omega\rightarrow0$, the power remains finite if $\sigma<2$, but is unbounded otherwise. 
It follows from Eq.~(\ref{eq:G_pnl}) that $G_1 < G_0 < G_1^{\rm cr}$, and therefore according to Table~\ref{tab:stab_symm} the corresponding localized modes are always stable.

\subsection{Mixed impurity with attractive nonlinearity} 
            \label{sect:pw_defoc_ri}

\begin{figure}
  \vspace*{-10mm}
  \setlength{\epsfxsize}{8.0cm}
  \centerline{\mbox{\epsffile{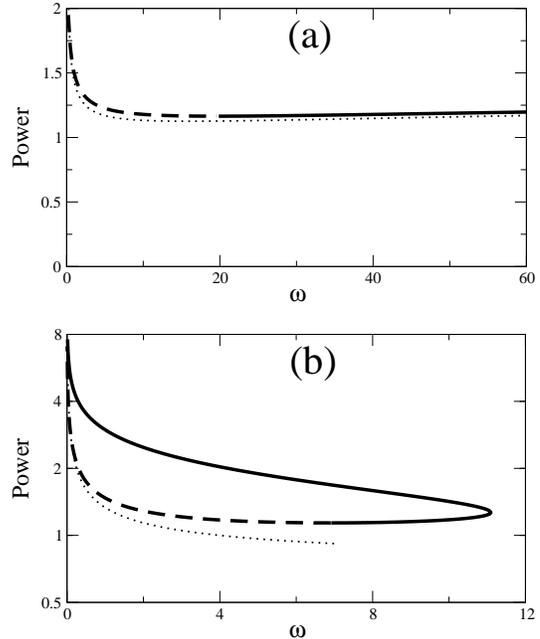}}}
  \vspace*{-10mm}
  \caption{ \label{fig:pwr-df-al_m-beta_p}
  Power vs.~frequency diagram for defocusing bulk media with $\alpha=-2$  and
  $\beta = 3$. (a) $\sigma = 1$, $\gamma = 0.8$; (b) $\sigma = 3$, $\gamma    = 1$, $\beta_0 \simeq 2.38$.
  Dotted curve: asymptotic given by the power $P_{\rm i}(\omega)$ of linear 
  impurity modes.}
\end{figure}

\begin{figure}
  \setlength{\epsfxsize}{7.0cm}
  \vspace*{-25mm}
  \centerline{\mbox{\epsffile{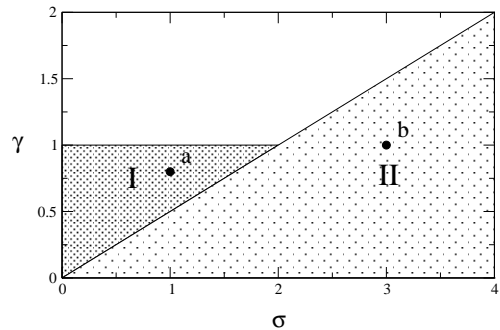}}}
  \vspace*{-25mm}
  \caption{ \label{fig:stab-df-al_m-beta_p}
  Stability regions for localized modes with $\alpha<0$ and $\beta>0$ in a 
  defocusing medium: 
  I: the modes are stable for frequencies above a certain threshold;
  II: localized modes can exist and be stable in a bounded 
      region of frequencies. 
  In the region without shading localized modes exist for all $\omega>0$, 
  but are unstable.
  The points (a) and (b) correspond to 
  Figs.~\ref{fig:pwr-df-al_m-beta_p}(a) and 
  \ref{fig:pwr-df-al_m-beta_p}(b), respectively. }
\end{figure}

Finally, we consider the case when localized modes do not exist in the 
linear limit. i.e.~when $\alpha<0$ and $\beta>0$. As was pointed out
above, the properties of localized modes depend on the relative strength 
of the bulk and defect nonlinearities, characterized by the ratio $\Gamma$. 
We have analyzed Eq.~(\ref{eq:p}) and found that in the case of a ``strong'' defect with $\Gamma>1/2$ the modes exist for all frequencies $\omega>0$, and correspond to a single branch in the $P(\omega)$ diagram, which asymptotically approaches $P_{\rm i}(\omega)$ at high frequencies, according to Eq.~(\ref{eq:P_inf}) [see Fig.~\ref{fig:pwr-df-al_m-beta_p}(a)]. Thus, these modes are stable above a certain cut-off, $\omega>\omega_s$, only if the impurity supports stable modes, i.e.~for $\gamma<1$. This parameter range corresponds to region~I in the diagram shown in Fig.~\ref{fig:stab-df-al_m-beta_p}.  Note that for $\omega\gg\omega_0$ the linear impurity response is negligible and the mode characteristics should not depend on the sign of $\alpha$.  Indeed, we see that region~I is the same in the cases shown in both Fig.~\ref{fig:stab-df-al_p-beta_p} and Fig.~\ref{fig:stab-df-al_m-beta_p}.

For $\Gamma<1/2$ (region II in Fig.~\ref{fig:stab-df-al_m-beta_p}) the 
modes only exist if the nonlinearity of the impurity exceeds the threshold 
value defined by Eq.~(\ref{eq:beta_0}), i.e.~for $\beta>\beta_0$. 
Then $P(\omega)$ has two branches, which appear at $\omega = 0$ and merge 
again at the critical point $\omega_s$, as shown in 
Fig.~\ref{fig:pwr-df-al_m-beta_p}(b). 
From the properties of the critical points discussed in Sec.~\ref{sect:stab} 
we conclude that the stable modes correspond to (i) the upper branch ($0<
\omega<\omega_s$) and (ii) part of the lower branch close to $\omega_s$, for 
which the slope $dP/d\omega$ is positive.

\section{Collapse dynamics} \label{sect:collapse}
\subsection{Virial relation} \label{sect:collapse_virial}

The nonlinearity-induced energy localization is a fundamental physical problem. The localization can occur in the form of stationary nonlinear impurity modes, and it is essential to understand the underlying physical mechanisms leading to localized states. A very efficient way for energy localization is the so-called {\em collapse (or ``blow-up'') dynamics} when, under certain conditions, nonlinear self-focusing dominates over diffraction, leading to an unlimited growth of the field intensity in a finite time. In real physical systems, the actual ``blow-up'' never occurs, however, the initial collapse dynamics can be correctly described in the framework of the continuous model equations, as long as the corresponding assumptions are not violated.

In this section we briefly discuss the collapse dynamics in the frame of our model equations~(\ref{eq:nls}),(\ref{eq:def_F}) and derive sufficient conditions for collapse in the presence of an impurity. For the sake of simplicity, we are considering the case of the power-law nonlinearities, as introduced in Sec.~\ref{sect:power_nl_full}.

Studying the collapse-induced effects due to a {\em nonlinear impurity}, we distinguish two cases: (i)~collapse in a bulk medium, away from the impurity site, and (ii)~collapse at the impurity site. In the first case, the impurity acts similar to a small perturbation, and thus collapse can occur if the power of the nonlinearity exceeds the critical value for a homogeneous self-focusing (i.e. $\rho=+1$) bulk medium, $\sigma_{\rm cr} = 2$~\cite{berge}. We note that, for the case of a nonlinearly repulsive impurity (e.g., $\beta<0$) at high intensities, the maximum field amplitude is always achieved away from the impurity site so that collapse can only occur in the bulk, see, e.g., Fig.~\ref{fig:bpm-sf-al_p-beta_m}(b). Collapse at the impurity site can only take place if the impurity possesses attractive nonlinearity (i.e. $\beta>0$). We hereafter consider this case.

In order to analyze the collapse conditions, we assume that the initial profile of the impurity mode is symmetric, and therefore $\psi(x) = \psi(-x)$ at all $t \ge 0$ due to the symmetry of Eq.~(\ref{eq:nls}). Then, the {\em effective mode width} $R$ can be defined as follows:
\begin{equation}
  R^2(t) = \frac{1}{P} \int_{-\infty}^{+\infty} x^2 |\psi(x,t)|^2 d x .
\end{equation}
We determine the temporal evolution of $R(t)$ by following the standard procedure~\cite{berge} and derive the so-called {\em virial relation}:
\begin{equation} \label{eq:virial}
 \begin{array}{l}
 { \displaystyle
   P \frac{d^2 ( R^2 ) }{d\; t^2} 
    = 8 H  + 4 \alpha I_0 +
 } \\*[9pt] { \displaystyle \quad
    4 \rho\; \frac{(2-\sigma)}{(1+\sigma)}
      \int_{-\infty}^{+\infty} |\psi|^{2\sigma+2} dx
     + 4 \beta\; \frac{(1-\gamma)}{(1+\gamma)} I_0^{\gamma+1} ,
 } \end{array}
\end{equation}
where $P$ and $H$ are the power and Hamiltonian defined by Eqs.~(\ref{eq:P}) and~(\ref{eq:H}), respectively, while $I_0 = |\psi (0, t)|^2 $ here designates the intensity at the impurity site. Because $P$ and $H$ are conserved quantities, Eq.~(\ref{eq:virial}) can be integrated when the powers of the nonlinearities attain the {\em critical values}: $\sigma_{\rm cr} = 2$, $\gamma_{\rm cr} = 1$, and linear defect response vanishes ($\alpha = 0$):
\begin{equation}
  R(t) = \sqrt{ R_0^2 + 4 t^2 (H / P) } ,
\end{equation}
where $R_0$ is the width of the input mode, and we assume that the initial mode profile does not have a phase modulation. If $H < 0$, it immediately follows that the mode width decreases and eventually vanishes at a finite time, indicating that the peak intensity goes to infinity since the power $P$ is conserved, thus the energy collapses to a single point~--- the impurity site. Therefore, a negative value of the Hamiltonian is a {\em sufficient condition} for collapse in this case.

If the power of the impurity nonlinearity is enhanced, i.e. $\gamma > \gamma_{\rm cr}$, the corresponding term on the right-hand side of Eq.~(\ref{eq:virial}) becomes negative, indicating an increase in the collapse growth rate. Increasing the power of bulk nonlinearity above the critical value, $\sigma > \sigma_{\rm cr}$ for $\rho=+1$, or decreasing the self-defocusing response, $\sigma < \sigma_{\rm cr}$ for $\rho=-1$, we observe a similar effect. In all these cases, the relation $H < 0$ remains a sufficient condition for collapse. The same argument holds for a defect with linearly repulsive response, $\alpha<0$. On the other hand, as we have demonstrated in the previous sections, a linearly attractive defect can support stable localized modes which do not exhibit collapse. In this case, and more generally when at least one of the last three terms on the right-hand side of Eq.~(\ref{eq:virial}) becomes positive, we can not directly use the virial relation to predict the collapse conditions.

\subsection{Collapse conditions} \label{sect:collapse_condition}

In order to predict whether collapse is possible for arbitrary values of the nonlinearity parameters we may employ the connection between the collapse dynamics and the properties of stationary localized modes. First, we note that for the stationary solutions the right-hand side of Eq.~(\ref{eq:virial}) is identical zero. For the critical case a perturbation of the solution that makes $H$ negative will result in a collapse. Thus, the collapse may occur when the high-frequency symmetric localized modes are unstable. Indeed, in the limit of high intensities we can neglect the linear impurity response by taking $\alpha =0$, and then an exact collapsing solution can be obtained (at $\sigma = 2$ and $\gamma=1$) in the form:
\begin{equation} \label{eq:slv_collapse}
 \begin{array}{l}
 { \displaystyle
   \psi(x, t) = \sqrt{\lambda(t)}\; u[\lambda(t) x] e^{-i \theta(x,t)} ,
 } \\*[9pt] { \displaystyle
    \theta(x,t) = \omega C \lambda(t) - x^2 {\lambda(t)} / {(4 C)} ,
 } \end{array}
\end{equation}
where $\lambda(t) = C (t - t_0)^{-1}$, $C$ is an arbitrary positive constant, $t_0$ is the collapse time, and $u(x)$ is the profile of a stationary localized mode with the frequency $\omega$. Although the solution~(\ref{eq:slv_collapse}) is unstable and therefore does not describe the actual collapse dynamics, it demonstrates a link between the stationary modes and the collapse phenomenon. Moreover, for a homogeneous self-focusing medium it was proved that an unstable soliton will collapse if its power is slightly increased (see, e.g., Ref.~\cite{sharp_collapse}). Then, we arrive at the following statement: {\em Collapse at the nonlinearly attractive impurity can be observed if and only if the stationary symmetric impurity modes exist in the limit of high frequencies and their power is bounded from above}, i.e. $P(\omega \rightarrow +\infty) < P_{\rm max}$, where $P_{\rm max}$ is a constant. We note, however, that the stationary modes themselves are not necessarily unstable in the case with the critical power of the impurity nonlinearity, as their power may asymptotically approach a constant from below (see the example in Fig.~\ref{fig:pwr-sf-al_p-beta_p_s1g1}). Finally, using the mode properties derived in Sec.~\ref{sect:power_nl_full}, we determine the parameter regions where collapse can occur, and summarize these results in Table~\ref{tab:collapse}.

\noindent \parbox[l]{8cm}{
\noindent
\begin{table}
\caption{ Collapse conditions }
\begin{tabular}{lll}
     Impurity response
   & Bulk nonlinearity
   & Collapse
\\ \hline \hline
     arbitrary
   & $\rho=+1$ and $\sigma \ge 2$
   & bulk
\\ \hline
     $\beta>0$ and $\gamma \ge 1$
   & $\rho=+1$
   & 
   \\
   & $\rho=-1$ and $\sigma \le 2 \gamma$
   & impurity
   \\
   $\beta >0$ and $\gamma < 1$
   & $\rho=+1$ and $\sigma \ge 2$
   &
\\ \hline
     $\beta < 0$ and/or $\gamma < 1$
   & $\rho=-1$ and/or $\sigma < 2$
   & does not occur
   \\
     $\beta > 0$
   & $\rho=-1$ and $\sigma > 2 \gamma$
   &
\end{tabular}
\label{tab:collapse}
\end{table}
}
\vspace*{-5mm}

\subsection{Collapse suppression} \label{sect:collapse_suppression}

\begin{figure}
\setlength{\epsfxsize}{11.0cm}
\vspace*{-10mm}
\centerline{\mbox{\epsffile{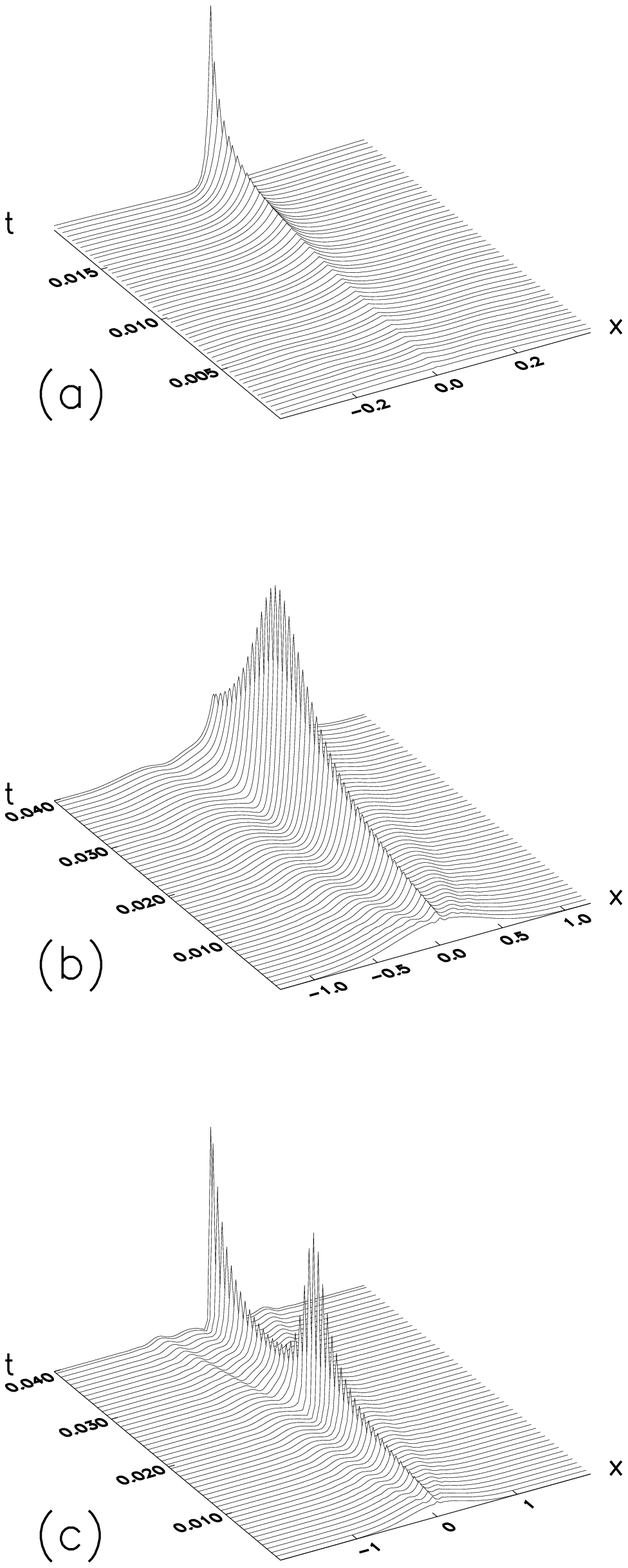}}}
\vspace{2mm}
\caption{ \label{fig:collapse}
Comparison of different types of the collapse-induced dynamics.
(a)~Impurity-induced collapse~--- the continuum model~(\ref{eq:nls}) for a
self-focusing Kerr medium ($\sigma=\gamma=1$,
$\alpha=2$, $\beta=0.55$);
(b)~Collapse suppression for a layer of a finite width ($d = 0.02$);
(c)~Collapse suppression due to the model discreteness ($h \simeq 0.02$). }
\end{figure}

The infinite growth of the field amplitude that results from the collapse dynamics would not occur in a real physical system. However, in the framework of the model~(\ref{eq:nls}),(\ref{eq:def_F}) we observe that collapse can be induced by an impurity even in a standard nonlinear Kerr medium, i.e., when $\sigma = \gamma = 1$, as demonstrated in Fig.~\ref{fig:collapse}(a). The final stage of self-focusing, preceding the ``blow-up'', should be described by modified equations that take into account some {\em additional physical effects which can no longer be neglected when the collapse is approached}. Below, we consider two possible mechanisms for such a {\em collapse suppression}.

The first example is self-focusing of spatial nonlinear guided optical modes supported by a thin nonlinear waveguide embedded in a bulk Kerr medium. When the field localization becomes very high, the delta-function can no longer be used to approximate the layer response, and then the nonlinear response should be modified as follows 
\begin{equation} \label{eq:opt_layer}
  {\cal F}(I; x) = \left\{ \begin{array}{l}
                 I,\, |x| > d/2 , \\
                 \widetilde{\alpha} + \widetilde{\beta} I,\, |x| \le d/2 ,
           \end{array} \right.
\end{equation}
where $d$ is the layer width, and the layer response parameters are related to those used in the delta-function approximation, $\widetilde{\alpha} d = \alpha$ and $\widetilde{\beta} d = \beta$. As can be seen from Fig.~\ref{fig:collapse}(b), a decrease of the mode width and a growth of the peak amplitude stop when the energy gets localized inside the nonlinear layer.

As the second example, we study the energy localization in an intrinsically discrete system such as a waveguide array. It has been demonstrated that in homogeneous lattices the collapse and the infinite growth of the peak amplitude is always suppressed~\cite{discrete_collapse}. This happens due to the presence of a minimal transverse scale, the characteristic distance~$h$ between the lattice sites or, for the present example, the width of the individual waveguides in the array. Therefore, when the width of the localized mode becomes comparable with $h$, the evolution can no longer be described in the framework of the continuum approximation. Therefore, the original model~(\ref{eq:nls}) should be modified to take the form of the discrete NLS-type equation:
\begin{equation} \label{eq:discrete}
     i \frac{\partial \psi_n}{\partial t}
     + \frac{(\psi_{n+1} + \psi_{n-1} - 2 \psi_n)}{ 2 h^2}
     + {\cal F}(I_n; n) \psi_n = 0,
\end{equation}
where $n$ is the site number. An example of the energy
localization in a discrete system is presented in
Fig.~\ref{fig:collapse}(c). When the mode becomes narrow, its
confinement is defined by the discreteness, and the final state
corresponds to breather-like oscillations near a stable
stationary solution of a discrete model.

\begin{figure}
\setlength{\epsfxsize}{8.0cm}
  \vspace*{-25mm}
\centerline{\mbox{\epsffile{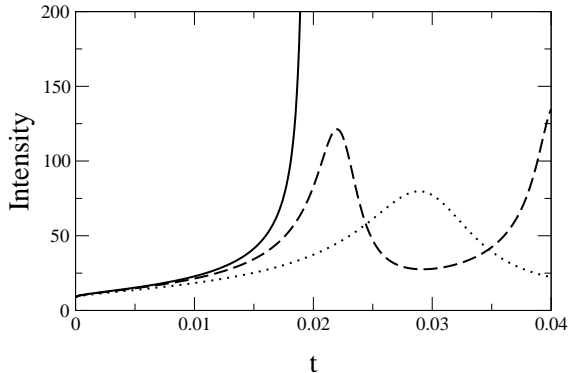}}}
  \vspace*{-25mm}
\caption{ \label{fig:collapse_intensity}
Evolution of the field intensity at the nonlinear impurity
corresponding to the plots shown in
Fig.~\ref{fig:collapse}(a)~--- solid, Fig.~\ref{fig:collapse}(b)~--- dotted, and Fig.~\ref{fig:collapse}(c)~--- dashed,
respectively.
}
\end{figure}

Finally, we note that the initial stage of the collapse-induced
dynamics is the same for all three different models, as is clearly
seen in Fig.~\ref{fig:collapse_intensity}. It is also important
to mention that the power of the high-frequency stationary
localized solutions of the modified
models~(\ref{eq:opt_layer}),(\ref{eq:discrete}) is {\em no longer
limited}, and thus the observation of collapse suppression
completely agrees with our general criterion introduced in
Sec.~\ref{sect:collapse_condition}.

\section{Conclusions}

We have analyzed spatially localized nonlinear modes supported by a point-like impurity, in the framework of the generalized nonlinear Schr\"odinger equation. We have considered three possible types of such nonlinear impurity modes, i.e. symmetric one- and two-hump modes and asymmetric one-hump mode, and described their regions of existence and stability, for both focusing and defocusing nonlinearity of a bulk medium and two different (attractive or repulsive) types of the impurity. In particular, we have obtained an analytical stability criterion for nonlinear localized modes based on the results of the linear stability analysis of the generalized NLS equation. For more specific physical applications, we have presented a detailed analysis of the nonlinear impurity modes and their stability in the case of the power-law nonlinearities in both the medium and defect, and discussed several scenarios of the instability-induced dynamics of the nonlinear impurity modes. In particular, we have described a novel physical mechanism of the energy localization due to the impurity-induced collapse of a nonlinear mode at the defect site, which can occur when the power of nonlinearity in the defect exceeds a critical value (i.e., $\gamma \ge 1$); this effect can be observed for the Kerr medium as well.

The problem we have analyzed above has a number of important physical applications ranging from the nonlinear dynamics of solids to the theory of nonlinear photonic crystals and waveguide arrays in nonlinear optics. In particular, our results can be linked to different special cases of the theory of nonlinear guided waves in layered dielectric media, and they provide also a generalization of the theory of nonlinear impurity modes in solids, together with a systematic classification of nonlinear impurity modes and the analysis of the mode stability and its instability-induced dynamics. Additionally, this problem can be considered as one of the first steps towards a deeper understanding of the competition between two different physical mechanisms of energy localization. In particular, we have presented our results emphasizing the cases where one can observe a clear evidence of competition between the disorder- and nonlinearity-induced localization.

\section*{Acknowledgments}

This work has been supported by the Australia-Denmark collaborative project of the Department of Industry, Science and Tourism, the Australian Photonics Cooperative Research Centre, and Planning and Performance Foundation grant. A.~A.~Sukhorukov is greatful for a warm hospitality of the Department of Mathematical Modelling at the Technical University of Denmark and a support of the Danish Graduate School in Nonlinear Science. O.~Bang acknowledges support from the Danish Technical Research Council through Talent Grant No.~9800400. The authors are thankful to Yuri Gaididei for useful suggestions.

\end{multicols}

\begin{references}

\bibitem{book} 
See, e.g., A.~A. Maradudin, 
{\em Theoretical and Experimental Aspects of the Effects of Point Defects 
and Disorder on the Vibrations of Crystal} 
(Academic Press, New York, 1966).

\bibitem{lifshitz}
See, e.g.,
I.~M. Lifshitz, Nuovo Cimento Suppl. {\bf 3}, 716 (1956); 
I.~M. Lifshitz and A.M. Kosevich, Rep. Progr. Phys. {\bf 29}, 217 (1966).

\bibitem{acoustic} 
See, e.g., A.~F. Asainov, K.~S. Len, and I.~Yu. Solodov, 
Akust. Zh. {\bf 39}, 592 (1993) [Sov. Phys. Acoust. {\bf 39}, 311 (1993)].

\bibitem{super} 
See, e.g., A.~F. Andreev, 
Pis'ma Zh. \'Eksp. Teor. Fiz. {\bf 46}, 463 (1987) 
[JETP Lett. {\bf 46}, 584 (1987)].

\bibitem{highTc}
A.~V.~Balatsky, Nature {\bf 403}, 717 (2000),
and references therein.

\bibitem{molina} 
M.~I. Molina and G.~P. Tsironis, 
Phys. Rev. B {\bf 47}, 15330 (1993);
D.~H. Dunlap, V.~M. Kenkre, and P. Reineker, 
Phys. Rev. B {\bf 47}, 14842 (1993).

\bibitem{molina2} 
G.~P. Tsironis, M.~I. Molina, and D. Hennig, 
Phys. Rev. E {\bf 50}, 2365 (1994).

\bibitem{souk} 
E. Lidorikis, K. Busch, Qiming Li, C.~T. Chan, and C.~M. Soukoulis, 
Phys. Rev. B {\bf 56}, 15090 (1997).

\bibitem{pc} 
S.~Y. Lin, E. Chow, V. Hietala, P.~R. Villeneuve, and J.~D. Joannopoulos, 
Science {\bf 282}, 274 (1998); 
M.~G. Khazhinsky and A.~R. McGurn, Phys. Lett. A {\bf 237}, 175 (1998);
A.~M. Zheltikov, S.~A. Magnitski\u{i}, and A.~V. Tarasishin, 
Pis'ma Zh. \'Eksp. Teor. Fiz. {\bf 70}, 323 (1999) 
[JETP Lett. {\bf 70}, 323 (1999)].

\bibitem{arrays}  
For the theory, see
W. Kr\'olikowski and Yu.~S. Kivshar, 
J. Opt. Soc. Am. B {\bf 13}, 876 (1996); 
and for experiments, see  
U. Peschel, R. Morandotti, J.~S. Aitchison, H.~S. Eisenberg, 
and Y. Silberberg, Appl. Phys. Lett. {\bf 75}, 1348 (1999).


\bibitem{flach}
S.~Flach and C.~R.~Willis, Phys. Rep. {\bf 295}, 181 (1998).

\bibitem{review} 
See, e.g., an earlier review paper, 
S.~A. Gredeskul and Yu.~S. Kivshar, Phys. Rep. {\bf 216}, 1 (1992).

\bibitem{books}  
{\em Disorder and Nonlinearity}, 
A.~R. Bishop, D.~K. Campbell, and S. Pnevmatikos, Eds.  
(Springer-Verlag, New York, 1989); 
{\em Disorder with Nonlinearity}, 
F. Abdullaev, A.~R. Bishop, and S. Pnevmatikos, Eds. 
(Springer-Verlag, Berlin, 1992).

\bibitem{tsir} 
See, e.g., 
V.~M. Kenkre and D.~K. Campbell, Phys. Rev. B {\bf 34}, 4959 (1986);
V.~M. Kenkre and G.~P. Tsironis, Phys. Rev. B {\bf 35}, 1473 (1987).

\bibitem{magnet} 
N.~N. Chen and M.~G. Cottam, Sol. State Commun. {\bf 84}, 379 (1992);
N.~N. Chen, M.~G. Cottam, and A.~F. Khater, 
Phys. Rev. B {\bf 51}, 1003 (1995);
S.~J. Xiong and S.~N. Evangelon, Physica D {\bf 81}, 111 (1995);
S. Rakhmanova and D.~L. Mills, Phys. Rev. B {\bf 58}, 11458 (1998).

\bibitem{optics} 
See, e.g., the pioneering papers: 
W.~J. Tomlinson, Opt. Lett. {\bf 5}, 323 (1980); 
N.~N. Akhmediev, Zh. \'Eksp. Teor. Fiz. {\bf 83}, 545 (1982)
[Sov. Phys. JETP {\bf 56}, 299 (1982)]; 
some review papers: 
G.~I. Stegeman, C.~T. Seaton, W.~H. Hetherington, A.~D. Boardman, 
and P. Egan, 
in {\em Electromagnetic Surface Excitations}, 
Eds. R.~F. Wallis and G.~I. Stegeman (Springer-Verlag, Berlin, 1986); 
F. Lederer, U. Langbein, and H.~E. Ponath, 
in {\em Lasers and Their Applications}, Ed. A.Y. Spassov 
(World Scientific, Singapore, 1987); 
more recent publications: 
K. Ogusu, J. Lightwave Tech. {\bf 8}, 1541 (1990);
R.~W. Micallef, Yu.~S. Kivshar, J. Love, D. Burak, and R. Binder, 
Opt. Quantum Electron. {\bf 30}, 751 (1998); 
and references therein.

\bibitem{ms} 
D.~J. Mitchell and A.~W. Snyder, 
J. Opt. Soc. Am. B {\bf 10}, 1572 (1993).

\bibitem{malomed} 
Yu.~S. Kivshar and B.~A. Malomed, 
J. Phys. A: Math. Gen. {\bf 21}, 1553 (1988).

\bibitem{nls_imp}
Yu.~S. Kivshar, Phys. Lett. A {\bf 161}, 80 (1991);
A.~D. Boardman, V. Bortolani, R.~F. Wallis, K. Xie, and H.~M. Mehta,
Phys. Rev. B {\bf 52}, 12736 (1995).

\bibitem{ltp} 
M.~M. Bogdan, A.~S. Kovalev, and I.~V. Gerasimchuk, 
Fiz. Nizk. Temp. {\bf 23}, 197 (1997) 
[Low Temp. Phys. {\bf 23}, 145 (1997)].

\bibitem{kosevich}
See, e.g., 
Yu.~S. Kivshar, A.~M. Kosevich, and O.~A. Chubykalo, 
Zh. \'Eksp. Teor. Fiz. {\bf 93}, 5968 (1987)
[Sov. Phys. JETP {\bf 66}, 545 (1987)]; 
Phys. Lett. A {\bf 125}, 35 (1987);
and references therein.

\bibitem{titchmarsh} 
E.~C. Titchmarsh, 
{\em Eigenfunction Expansions Associated with 
Second-Order Differential Equations} 
(Oxford University Press, London, 1958).

\bibitem{vk} 
N.~G. Vakhitov and A.~A. Kolokolov, 
Izv. Vyssh. Uchebn. Zaved. Radiofiz. {\bf 16}, 1020 (1973) 
[Radiophys. Quant. Electron. {\bf 16}, 783 (1973)].

\bibitem{kuznetsov} 
E.~A. Kuznetsov, A.~M. Rubenchik, and V.~E. Zakharov, 
Phys. Rep. {\bf 142}, 103 (1986).

\bibitem{jones} 
C.~K.~R.~T. Jones, 
Ergodic Theory and Dynamical Systems {\bf 8$^{\ast}$}, 119 (1988).  

\bibitem{grillakis} 
M. Grillakis, J. Shatah, and W. Strauss,
J. Func. Anal. {\bf 74}, 160 (1987); {\em ibid} {\bf 94}, 308 (1990).

\bibitem{berge}
J.~J. Rasmussen and K. Rypdal, Physica Scripta {\bf 33}, 481 (1986);
see also other reviews
E.~A. Kuznetsov, Chaos {\bf 6}, 381 (1996);
L. Berg\'e,  Phys. Rep. {\bf 303}, 259 (1998);
see also C. Sulem and P.-L. Sulem,  
{\em The Nonlinear Schr\"odinger Equation: Self-Focusing and Wave Collapse} (Springer-Verlag, New York, 1999).

\bibitem{pelin}
D.~E.~Pelinovsky, V.~V.~Afanasjev, and Yu.~S.~Kivshar, 
Phys. Rev. E {\bf 53}, 1940 (1996).

\bibitem{sharp_collapse}
S.~K. Turitsyn, Phys. Rev. E {\bf 47}, R13 (1993); 
E.~A. Kuznetsov, J.~J. Rasmussen, K. Rypdal, and S.~K. Turitsyn,
Physica D {\bf 87}, 273 (1995).

\bibitem{discrete_collapse}
O. Bang, J.~J. Rasmussen, and P.~L. Christiansen,
Nonlinearity {\bf 7}, 205 (1994);
E.~W. Laedke, K.~H. Spatschek, and S.~K. Turitsyn,
Phys. Rev. Lett {\bf 73}, 1055 (1994);
B. Malomed and M.~J. Weinstein,
Phys. Lett. A {\bf 220}, 91 (1996).


\end{references}
\end{document}